\documentclass{article}

\usepackage[english]{babel}

\usepackage[letterpaper,top=2cm,bottom=2cm,left=2cm,right=2cm,marginparwidth=1.75cm]{geometry}

\usepackage{mathrsfs}
\usepackage{amsmath}
\usepackage{amssymb}
\usepackage{epsfig}
\usepackage{graphicx}
\usepackage{subfigure}
\usepackage{multirow}
\usepackage{epstopdf}
\usepackage{authblk}
\usepackage{cite}

\def\be{\begin{eqnarray}}
\def\en{\end{eqnarray}}
\def\non{\nonumber\\}

\title{Quasi-two-body decays $B_c \rightarrow \ K^{*} h \rightarrow K \pi h $ in the perturbative QCD}
\author{   Zi-Yu Zhang$^1$, Zhi-Qing Zhang$^1$\footnote{Electronic address: zhangzhiqing@haut.edu.cn (corresponding author)}, Si-Yang Wang$^2$, Zhi-Jie Sun$^1$ and You-Ya Yang$^1$} 
\affil{ \it \small $^1$  Institute of Theoretical Physics, School of Sciences, Henan University
of Technology,\\ \it Zhengzhou, Henan 450052, China; \\
\small $^2$ \it
Institute of Particle Physics and Key Laboratory of Quark and Lepton Physics (MOE),\\
\small \it Central China Normal University, Wuhan, Hubei 430079, China }
\date{\today}

\begin{document}
\maketitle

\begin{abstract}
In this work we study the quasi-two-body decays $B_c \rightarrow \ K^{*} h \rightarrow K \pi h (h = D, D_s, K, \pi, \eta, \eta')$  in the perturbative QCD (PQCD) approach. The two-meson distribution amplitudes (DAs) $\Phi^{\text{P-wave}}_{K\pi}$ are introduced to describe the final state interactions of the $K \pi$ pair, which involve the time-like form factors $F_{K\pi}(s)$ parameterized by the relativistic Breit-Wigner function and the Gegenbauer polynomials. We calculate the branching ratios for these quasi-two-body decays, from which one can obtain the branching raios for the corresponding two-body decays under the narrow width approximation relation. We find that
$B^+_c\to K^{*+}D^0$ and $B^+_c\to K^{*0}D^+$ have the largest branching ratios, which can reach up to $10^{-6}$, while the branching ratios for other two-body decays are very small and only about $10^{-8}\sim10^{-7}$. As we expected that
 the branching ratios of the pure annihilation decays are usually small, while in our considered such type of decays, the channel $B_c^+ \rightarrow \bar K^{*0}\ K^{+}$ has the largest branching ratio, which is near $10^{-6}$. These results are consistent with the previously PQCD calculations obtained in the two-body framework, which can be tested by the future LHCb
 experiments. For the decays $B_c^+ \rightarrow K^{*+} D^{0}\to K^{0}\pi^+D^{0}$ , $B_c^+ \to K^{*0}D^{+}\to \ K^{+}\pi^-D^{+}$ and $B_c^+ \to \bar K^{*0}D_s^{+}\to K^{-}\pi^+D_s^{+}$, we calculate their direct CP violations and find that $A_{CP}(B_c^+ \to K^{*+}D^{0}\to K^{0}\pi^+D^{0})=(-14.6_{-1.12}^{+9.19})\% $ is the largest one, which is possible measured by the present LHCb experiments. For the pure annihilation type decays, there is no CP violations because only the tree operators are involved. Furthermore, we also give the differential distributions of the branching ratios and the direct CP violations in the the $K\pi$ invariant mass $\omega$ for the decays $B_c\to K^* D_{(s)}\to K \pi D_{(s)}$.
\end{abstract}

{\centering\section{INTRODUCTION}\label{intro}}

In $B_{u,d,s,c}$ meson systems, the $B_c$ meson is the only quark-antiquark bound state $(\bar{b} c)$ composed of both heavy quarks with different flavors, and is thus flavor asymmetric. It can decay only via weak interaction, since the two flavor asymmetric quarks (b and c) can not be annihilated into gluons (photons) via strong (electromagnetic) interaction. Because each of the two heavy quarks can decay individually ($b\to c(u), c\to s(d)$ transitions), and they can also annihilate through weak interaction, the $B_c$ meson has many rich decay channels, and provides an ideal platform to study the nonleptonic weak decays of heavy mesons, to test the standard model, and to search for any new physics signals.
In recent years, some experimental studies on $B_c$ meson decays to three-body even more multihardon final states, such as $B_c\to K^+K^-\pi^+$ \cite{lhcb1}, $B_c\to J/\psi D^{(*)0}K^{+}, J/\psi D^{(*)+}K^{*0}$ \cite{lhcb2}, $B_c\to p\bar p \pi^+$ \cite{lhcb3}, $B_c\to J/\psi\pi^+\pi^-\pi^+, J/\psi K^+\pi^-\pi^+$, $J/\psi K^+ K^- K^+$\cite{lhcb4}, have been performed except for the $B_c$ two-body decays. These kinds of decays are getting more and more attentions, which are caused by the following reasons. These decays involve much more complicated QCD dynamics compared with two-body decay cases, because of entangled nonresonant
and resonant contributions, and significant final-state interactions. Many new resonance states are observed in the invariant mass distributions of the
multihardon final state decays, which are difficult to understand in terms of a common meson or baryon, and called as exotic states.
Furthermore, the direct CP violations for some of these decays can be analysed both in two-body and multi-body frameworks. In the muti-body framework, the direct CP asymmetry may depend on the invariant mass distributions of the meson-pair decaying from some internal resonances, and is (strongly) affected by the finite widths of the resonances. While in the
two-body framework with the resonance masses being fixed, the direct CP violation is just a number, which may be overestimated or underestimated compared with the data. So it is important and necessary to research the $B_c$ meson decays to the three-body even more multihardon final states. In this work, one of our main purposes
is to check the width effect of the resonance state $K^*$ on the branching ratios and the direct CP violations for the quasi-two-body
decays $B_c \to K^{*} h \to K \pi h (h = D, D_s, K, \pi, \eta, \eta')$  in the PQCD approach. The other purpose
is to understand the annihilation contributions under the three-body framework. As a feature of the PQCD approach, the
annihilation type Feynman diagrams are calculable, which are important for the decays occurring through the weak annihilation diagrams only.

In fact, in order to study these quasi-two-body $B_{(c)}$ meson decays, many approaches based on symmetry principles and factorization theorems have been proposed. Symmetry principles incude the U-spin \cite{Bhat,Gronau,xu}, isospin and flavor $SU(3)$ symmetry \cite{Gronau2,Engelhard,Imbeault,He}, and the factorization-assisted topological-diagram amplitude (FAT) approach \cite{Zhou}, etc. Factorization theorems include the QCD-improved factorization approach \cite{Krankl,Cheng,Li,Cheng2007,Klein} and the PQCD approach \cite{zhao, zhang, zhang2, Liy, maaj}. It has been proposed that the factorization theorem of the quasi-two-body $B_{(c)}$ decays is approximately valid when the two particles move collinearly and the bachelor particle recoils back in the final states. According to this quasi-two-body-decay mechanism, the two-hadron distribution amplitudes (DAs) are introduced into the PQCD approach, where the strong dynamics between the two final hadrons in the resonant regions are included.

This paper is organized as follows. The framework of the PQCD approach for the quasi-two-body $B_{c}$ decays is reviewed in Section II,
where the kinematic variables for each meson are defined and the P-wave $K\pi$ pair distribution amplitudes up to twist-3 are parametrized.  Then, the analytical formulas of each Feynman diagram and the total amplitudes for these decays are listed.
In Section III, the numerical results and discussions are presented. The final section is devoted to our conclusions. Some
details related functions are collected in the Appendix.

{\centering\section{THE FRAMEWORK }}

In the framework of the PQCD approach for the quasi-two-body decays, the factorization formulas for the $B_c \rightarrow \ K^{*} h \rightarrow K \pi h$
decay amplitudes can be written as \cite{Chen:2002th,Chen:2004az}
\be
\mathcal{A}=\Phi_{B_c} \otimes H \otimes \Phi^{\text{P-wave}}_{K\pi} \otimes \Phi_{h},
\en
where $\Phi_{B_c}(\Phi_{h})$ denotes the DAs of the initial (final bachelor) meson, $\Phi^{\text{P-wave}}_{K\pi}$ is the P-wave $K\pi$ pair DAs, and
$\otimes$ denotes the convolution integrations over the parton momenta. Similar to the two-body decay case, the evolution of the hard kernel H for the b quark decay is calculable perturbatively and starts with the diagrams of single hard gluon exchange. The nonperturbative dynamics are absorbed into those DAs $\Phi_{B_c}, \Phi_{h}$ and $\Phi^{\text{P-wave}}_{K\pi}$.

In the rest frame of the $B_c$ meson, we define the $B_c$ meson momentum $p_{B_c}$, the $K (\pi)$ meson momentum $p_1 (p_2)$, the $K^*$ meson momentum $p=p_1+p_2$, and the bachelor meson $h$ momentum $p_3$ in light-cone coordinates as
\be
   p_{B_c}&=&\frac{m_{B_c}}{\sqrt{2}}\left(1,1, \mathbf{0}_{\mathrm{T}}\right), \quad p=\frac{m_{B_c}}{\sqrt{2}}\left(1-r^2, \eta, \mathbf{0}_\mathrm{T}\right), \quad p_3=\frac{m_{B_c}}{\sqrt{2}}\left(r^2, 1-\eta, \mathbf{0}_\mathrm{T}\right), \\
   p_1&=&\frac{m_{B_c}}{\sqrt{2}}\left(\zeta\left(1-r^2\right),(1-\zeta) \eta, \mathbf{p}_{1\mathrm{T}}\right), \quad p_2=\frac{m_{B_c}}{\sqrt{2}}\left((1-\zeta)\left(1-r^2\right), \zeta \eta, \mathbf{p}_{2\mathrm{T}}\right), \label{p12}
\en
where $\eta=w^2 /\left[\left(1-r^2\right) m_{B_c}^2\right]$ with the mass ratio $r=m_h / m_{B_c}$ , $\zeta$ is the momentum fraction for the kaon meson. The momenta of the light quarks in the $B_c$ meson, the $K^*$ meson and the bachelor meson $h$ are defined as $k_B, k$ and $k_3$, respectively
\begin{equation}
    k_B=\left(0,x_B p^-_B, \mathbf{k}_{B\mathrm{T}}\right), \quad k=\left(z p^{+}, 0, \mathbf{k}_{\mathrm{T}}\right), \quad k_3=\left(0, x_3 p_3^{-}, \mathbf{k}_{3\mathrm{T}}\right)
\end{equation}
where $x_B$, $z$ and $x_3$ are the corresponding momentum fractions .

{\centering\subsection{WAVE FUNCTIONS }}

In the course of the PQCD calculations, the necessary inputs contain the DAs, which are constructed via the nonlocal matrix elements. The $B_c$ meson light-cone matrix element can be decomposed as

\begin{equation}
    \int d^4 z e^{i k_1 \cdot z}\left\langle 0\left|\bar{b}_\alpha(0) c_\beta(z)\right| B_c\left(p_{B_c}\right)\right\rangle=\frac{i}{\sqrt{2 N_c}}\left[\left(p\hspace{-1.7truemm}/_{B_c}+M\right) \gamma_5 \phi_{B_c}\left(k_1\right)\right]_{\beta \alpha},
\end{equation}
where $N_c=3$ is the color factor. Here, we only consider the contribution from the dominant Lorentz structure. In coordinate space the distribution amplitude $\phi_{B_c}$ with an intrinsic $b$ (the conjugate space coordinate to $k_T$ ) dependence is adopted in a Gaussian form \cite{Liu:2018kuo}
\begin{equation}
    \phi_{B_c}(x, b)=N_{B_c} x(1-x) \exp \left[-\frac{(1-x) m_c^2+x m_b^2}{8 \omega_b^2 x(1-x)}-2 \omega_b^2 b^2 x(1-x)\right],
\end{equation}
where the shape parameter  $\omega_b = 1.0 \pm 0.1 \mathrm{GeV}$  related to the factor  $N_{B_{c}}$  by the normalization
$\int_{0}^{1} \phi_{B_{c}}(x, 0) d x=1$.

For D meson, the two-parton light-cone distribution amplitudes (LCDAs) in the heavy quark limit can be written as \cite{kuri,rhli}
\begin{equation}
    \left\langle D\left(p_3\right)\left|q_\alpha(z) \bar{c}_\beta(0)\right| 0\right\rangle=\frac{i}{\sqrt{2N_c}} \int_0^1 d x e^{i x p_3 \cdot z}\left[\gamma_5\left(p\hspace{-1.5truemm}/ _3+m_D\right) \phi_D(x, b)\right]_{\alpha \beta},
\end{equation}
with the distribution amplitude $\phi_D(x, b)$,
\begin{equation}
    \phi_D(x, b)=\frac{1}{2 \sqrt{2N_c}} f_D 6 x(1-x)\left[1+C_D(1-2 x)\right] \exp \left[\frac{-\omega^2 b^2}{2}\right],
\end{equation}
where $C_D=0.5 \pm 0.1, \omega=0.1 \mathrm{GeV}$ and $f_D=211.9 \mathrm{MeV}$. It is similar for the LCDAs of $D_s$ meson but with different parameters $C_{D_s}=0.4 \pm 0.1, \omega=0.2 \mathrm{GeV}$ and $f_{D_s}=249 \mathrm{MeV}$, caused by a
little SU(3) breaking effect \cite{Ma:2017kec}. As the LCDAs of the light pseudoscalar mesons $\pi, K, \eta^{(\prime)}$ up to twist-3 can be found in our recent work \cite{zhao}.

The P-wave $K\pi$ pair distribution amplitudes are defined as \cite{Ma:2019qlm}
\begin{equation}
\Phi_{K \pi}^{P \text {-wave }}=\frac{1}{\sqrt{2 N_c}}\left[p\hspace{-1.5truemm}/ \phi_0\left(z, \zeta, \omega^2\right)+\omega \phi_s\left(z, \zeta, \omega^2\right)+\frac{p\hspace{-1.5truemm}/_1 p\hspace{-1.5truemm}/_2- p\hspace{-1.5truemm}/_2 p\hspace{-1.5truemm}/_1}{\omega(2 \zeta-1)} \phi_t\left(z, \zeta, \omega^2\right)\right],
\end{equation}
with the functions
\begin{equation}
\begin{aligned}
\phi_0 & =\frac{3 F_{K \pi}(s)}{\sqrt{2 N_c}} z(1-z)\left[1+a_{1 K^*}^{\|} 3(2 z-1)+a_{2 K^*}^{\|} \frac{3}{2}\left(5(2 z-1)^2-1\right)\right] P_1(2 \zeta-1), \\
\phi_s & =\frac{3 F_s(s)}{2 \sqrt{2 N_c}}(1-2 z) P_1(2 \zeta-1), \\
\phi_t & =\frac{3 F_t(s)}{2 \sqrt{2 N_c}}(2 z-1)^2 P_1(2 \zeta-1),
\end{aligned}
\end{equation}
where the Legendre polynomial $P_1(2\zeta-1)=2\zeta-1$ and the Gegenbauer moments $ a_{1 K^*}^{\parallel}=0.05 \pm 0.02, a_{2 K^*}^{\parallel}=0.15 \pm 0.05$ \cite{Liy}. It is well known that the relativistic Breit-Wigner (RBW) function is an appropriate model for the narrow resonances which can be well separated from any other resonant or nonresonant contributions
with the same spin, and it is widely used in the experimental data analyses. Here the time-like form factor $F_{K \pi}(s)$ with $s=\omega^2=m^2(K\pi)$ via the RBW line shape \cite{aaij,rbw2,rbw3}
\begin{equation}
F_{K \pi}\left(s\right)=\frac{m_{K^*}^2}{m_{K^*}^2-s-i m_{K^*} \Gamma\left(s\right)},
\end{equation}
where the invariant mass-dependent width $\Gamma(s)$ is defined as
\begin{equation}
    \Gamma(s)=\Gamma_{K^*}\frac{m_{K^*}}{\sqrt{s}}\left(\frac{\left|\overrightarrow{p_1}\right|}{\left|\overrightarrow{p_0}\right|}\right)^{3} \frac{1+\left(\left|\overrightarrow{p_0}\right| r_{B W}\right)^2}{1+\left(\left|\overrightarrow{p_1}\right| r_{B W}\right)^2}.
\end{equation}
Here $|\overrightarrow{p}_1|$ is the magnitude of the momentum for the daughter meson $K$ or $\pi$ in the $K^*$ meson rest frame defined in the next subsection,
and $|\overrightarrow{p_0}|$ is the value of $|\overrightarrow{p_1}|$ at $s = m_{K^{*}}^2$.  The barrier radius $r_{BW} = 4.0 GeV^{-1}$ is taken as in Refs.\cite{aaij,rbw2,rbw3}.

{\centering\subsection{ Analytic formulae } \label{formu}}

For the quasi-two-body decays $B_c \rightarrow \ K^{*} D_{(s)} \rightarrow K \pi D_{(s)}$, the effective Hamiltonian relevant to the $b \to s(d) $ transition is given by \cite{heff}
\begin{equation}
\begin{aligned}
H_{e f f}= & \frac{G_F}{\sqrt{2}}\left\{\sum_{q=u, c} V_{q b} V_{q s(d)}^*\left[C_1(\mu) O_1^{(q)}(\mu)+C_2(\mu) O_2^{(q)}(\mu)\right]\right. \\
& \left.-\sum_{i=3 \sim 10} V_{t b} V_{t s(d)}^* C_i(\mu) O_i(\mu)\right\}+H . c .,
\end{aligned}
\end{equation}
where the Fermi coupling constant $G_F \simeq 1.166 \times 10^{-5} \mathrm{GeV}{ }^{-2}$, $V_{q b} V_{q d(s)}^*$ and $V_{t b} V_{t d(s)}^*$ are the products of the Cabibbo-Kobayashi-Maskawa (CKM) matrix elements. The scale $\mu$ separates the effective Hamiltonian into two distinct parts: the Wilson coefficients $C_i$, and the local four-quark operators $O_i$. The local four-quark operators for $b\to d$ transition are written as

\begin{figure}[htbp]
\centering
        \includegraphics[width=1\textwidth]{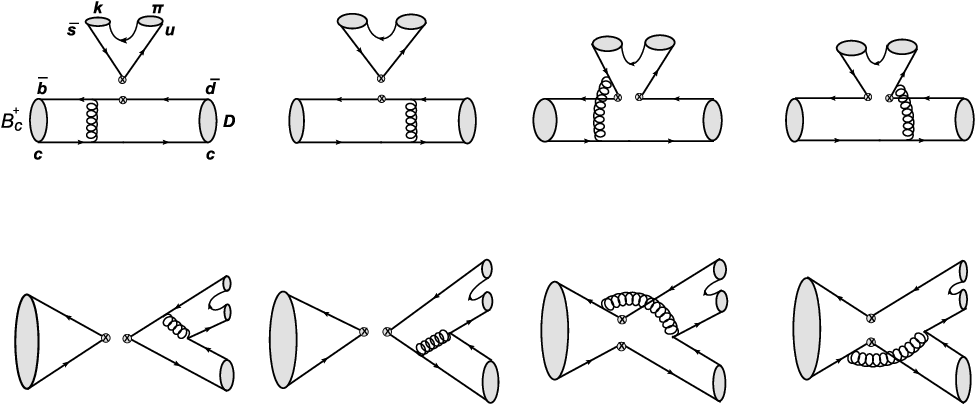}

          \caption{The leading order Feynman diagrams for the decay $B_c^+ \rightarrow K^{*+}D^{0} \rightarrow K^{0}\pi^+ D^{0}$. }
        \label{feynman}
   \end{figure}
\begin{equation}
\begin{aligned}
    O_1^{(q)} & =\left(\bar{d}_i q_j\right)_{V-A}\left(\bar{q}_j b_i\right)_{V-A}, & O_2^{(q)} &=\left(\bar{d}_i q_i\right)_{V-A}\left(\bar{q}_j b_j\right)_{V-A}, \\
O_3 & =\left(\bar{d}_i b_i\right)_{V-A} \sum_q\left(\bar{q}_j q_j\right)_{V-A}, & O_4 &=\left(\bar{d}_i b_j\right)_{V-A} \sum_q\left(\bar{q}_j q_i\right)_{V-A}, \\
O_5 & =\left(\bar{d}_i b_i\right)_{V-A} \sum_q\left(\bar{q}_j q_j\right)_{V+A}, & O_6 &=\left(\bar{d}_i b_j\right)_{V-A} \sum_q\left(\bar{q}_j q_i\right)_{V+A}, \\
O_7 & =\frac{3}{2}\left(\bar{d}_i b_i\right)_{V-A} \sum_q e_q\left(\bar{q}_j q_j\right)_{V+A}, & O_8 &=\frac{3}{2}\left(\bar{d}_i b_j\right)_{V-A} \sum_q e_q\left(\bar{q}_j q_i\right)_{V+A}, \\
O_9 & =\frac{3}{2}\left(\bar{d}_i b_i\right)_{V-A} \sum_q e_q\left(\bar{q}_j q_j\right)_{V-A}, & O_{10} &=\frac{3}{2}\left(\bar{d}_i b_j\right)_{V-A} \sum_q e_q\left(\bar{q}_j q_i\right)_{V-A},
\label{operator}
\end{aligned}
\end{equation}
where the color indices $\mathrm{i}$ and $\mathrm{j}$. Here $V \pm A$ refer to the Lorentz structures $\gamma_\mu\left(1 \pm \gamma_5\right)$. The
local four-quark operators for $b\to s$ transition can be obtained by replacing $ d$ with $s$ in Eq. (\ref{operator}). While for the pure annihilation decays $B_c \rightarrow \ K^{*} \pi(K, \eta^{(\prime)}) \rightarrow K \pi \pi(K, \eta^{(\prime)})$, the related weak effective Hamiltionian is given as
\be
H_{e f f}=  \frac{G_F}{\sqrt{2}} V_{c b} V_{us(d)}^*\left[C_1(\mu) O_1^{(q)}(\mu)+C_2(\mu) O_2^{(q)}(\mu)\right],
\en
with the single tree operators,
\be
O_1&=&\left(\bar{d}_i u_j\right)_{V-A}\left(\bar{c}_j b_i\right)_{V-A}, \; O_2 =\left(\bar{d}_iu_i\right)_{V-A}\left(\bar{c}_j b_j\right)_{V-A} \;\;\;(\text{for $b\to d$ transtion}),\\
O_1&=&\left(\bar{s}_i u_j\right)_{V-A}\left(\bar{c}_j b_i\right)_{V-A},\; O_2 =\left(\bar{s}_iu_i\right)_{V-A}\left(\bar{c}_j b_j\right)_{V-A} \;\;\;(\text{for $b\to s$ transtion}).
\en
The typical Feynman diagrams at the leading order for the quasi-two-body
decays $B_c \rightarrow \ K^{*} h \rightarrow K \pi h$ are shown in Fig. \ref{feynman}, where we take the decay
$B_c^+\to K^{*+}D^{0} \to K^{0}\pi^{+} D^{0}$ as an example.
We mark LL, LR, SP to denote the contributions from $(V-A)(V-A),(V-A)(V+A)$ and $(S-P)(S+P)$ operators, respectively. The amplitudes from the factorizable emission diagrams Fig. 1(a) and 1(b) are given as
\be
      \mathcal{F}_e^{L L}&=& 8 \pi \ F_{K \pi}m_{B_c}^4 C_F \int_0^1 d x_B d x_3 \int_0^{\infty} b_1 b_3 d b_1 d b_3 \phi_B\left(x_B, b_1\right) \phi_D\left(x_3, b_3\right) \non
      &&\times\left\{ \left[-\bar\eta\left[\eta(1-x_3)+x_3(1-2r)-2r_b\right]-rr_b(1+\eta)+r^2(x_3-2r_b)\right]\right.\non
      &&\left.\times\alpha_s (t_a) h\left(\alpha_e, \beta_a, b_1, b_3\right)  \exp \left[-S_{a b}\left(t_a\right)\right]S_t(x_3)\right. \non
      &&\left.+[\bar\eta(2r+\eta x_B)-2rx_B(1+\eta)+r^2(x_B-1)]\right.\non
      &&\left.\times\alpha_s (t_b) h\left(\alpha_e, \beta_b, b_1, b_3\right)  \exp \left[-S_{a b}\left(t_b\right)\right]S_t(x_B)\right\},\\
      \mathcal{F}_e^{LL}&=& \mathcal{F}_e^{LR}, \mathcal{F}_e^{SP}=0,
\en
Where $\bar\eta=1-\eta$ and the mass ratio $r_b=m_b/m_{B_c}$. $\eta$ and $r$ has been defined under Eq.(\ref{p12}). The hard function $h\left(\alpha_e, \beta_a, b_1, b_3\right)$, the hard scales $t_{a,b}$, the Sudakov factor $\exp \left[-S_{a b}(t)\right]$ and the threshold resummation factor $S_t(x)$ are given in Appendix A.
The amplitudes for the nonfactorizable eimission diagrams Fig. 1(c) and 1(d) are written as
\be
      \mathcal{M}_e^{L L}&= & 16 \sqrt{\frac{2}{3}} \pi m_{B_c}^4 C_F \int_0^1 d x_B d z d x_3 \int_0^{\infty} b_1 b d b_1 d b \phi_B\left(x_B, b_1\right) \phi_D\left(x_3, b_1\right) \phi_0\non
      &&\times\left\{[r[(1+\eta)(x_B-1)+\bar\eta x_3+\eta z]+(1-\eta^2)(1-x_B-z)]\right.\non
      &&\left.\times\alpha_s (t_c)h\left(\beta_c, \alpha_e,  b_1, b\right)  \exp \left[-S_{c d}\left(t_c\right)\right] \right.\non
      &&\left. +\left[(r-\bar\eta)(1-x_3)\bar\eta-r[(1+\eta)x_B-\eta z]+\bar\eta(2x_B-z)\right]\right.\non
      &&\left.\times\alpha_s (t_d)h\left(\beta_d, \alpha_e, b_1, b\right)  \exp \left[-S_{c d}(t_d)\right]\right\}, \\
      \mathcal{M}_e^{L R}&= & 16 \sqrt{\frac{2}{3}} \pi m_{B_c}^4 C_F \sqrt{\eta(1- r^2)} \int_0^1 d x_B d z d x_3 \int_0^{\infty} b_1 b d b_1 d b \phi_B\left(x_B, b_1\right)\phi_D\left(x_3, b_1\right)\phi_0  \non
      &&\times \left\{\left[\bar\eta(1-x_B-z)(\phi^s+\phi^t)-r(\bar\eta x_3-z)(\phi^s-\phi^t)+2r(1-x_B-z)\phi^s\right]\right.\non
      &&\left.\times\alpha_s (t_c)  h\left(\beta_c, \alpha_e, b_1, b\right)  \exp \left[-S_{c d}\left(t_c\right)\right]\right. \non
      &&\left.+\left[r[\bar\eta(x_3-1)+z](\phi^s+\phi^t)+2r(x_B-z)\phi^s+\bar\eta(x_B-z)(\phi^s-\phi^t)\right]\right.\non
      &&\left.\times\alpha_s (t_d)  h\left(\beta_d, \alpha_e, b_1, b\right)  \exp \left[-S_{c d}\left(t_d\right)\right]\right\}, \\
      \mathcal{M}_e^{SP}&= & 16 \sqrt{\frac{2}{3}} \pi m_{B_c}^4 C_F \int_0^1 d x_B d z d x_3 \int_0^{\infty} b_1 b d b_1 d b \phi_B\left(x_B, b_1\right) \phi_D\left(x_3, b_1\right) \phi_0\non
      &&\times\left\{\left[(r-\bar\eta)\bar\eta x_3-r(1+\eta)(1-x_B)-rz\eta+\bar\eta[2(1-x_B)-z]\right]\right.\non
      &&\left.\times\alpha_s (t_c)h\left(\beta_c, \alpha_e, b_1, b\right)  \exp \left[-S_{c d}\left(t_c\right)\right]\right. \non
      && \left.+\left[r[\bar\eta(1-x_3)+\eta(z-x_B)-x_B]+(1-\eta^2)(x_B-z)\right]\right.\non
      &&\left.\times\alpha_s (t_d)h\left(\beta_d, \alpha_e, b_1, b\right)  \exp \left[-S_{c d}\left(t_d\right)\right]\right\}.
\en
It is noticed that the integration of $b_3$ has been performed using $\delta$ function $\delta(b_1-b_3)$, leaving only integration of $b_1$ and $b_2$.
The amplitudes from the nonfactorizable annihilation diagrams Fig. 1(e) and 1(f) are listed as
\be
      \mathcal{M}_a^{L L}&=& 16 \sqrt{\frac{2}{3}} \pi m_{B_c}^4 C_F  \int_0^1 d x_B d z d x_3 \int_0^{\infty} b_1 b_3 d b_1 d b_3 \phi_B\left(x_B, b_1\right) \phi_D\left(x_3, b_3\right)\phi_0\non
      &&\times\left\{\left[\bar\eta[(1+\eta)(1-x_B-z)-r_b]\phi^0-r\sqrt{\eta(1-r^2)}[z(\phi^s+\phi^t)-\bar\eta x_3(\phi^s-\phi^t)]\right.\right.\non
      &&\left.\left.+2(2r_b+x_B-1)\phi^s\right]\alpha_s (t_e) h\left(\beta_e, \alpha_a, b_1, b_3\right)  \exp \left[-S_{ef}\left(t_e\right)\right] \right.\non
      &&\left.+\left[\bar\eta[\bar\eta x_3-x_B(1+\eta)+r_c+\eta z]\phi^0+r\sqrt{\eta(1-r^2)}\left[\bar\eta x_3(\phi^s+\phi^t)+z(\phi^s-\phi^t)\right.\right.\right.\non
      &&\left.\left.\left.+2(2r_c-x_B)\phi^s\right]\right]\alpha_s (t_f) h\left(\beta_f, \alpha_a, b_1, b_3\right)  \exp \left[-S_{ef}\left(t_f\right)\right]\right\},\\
      \mathcal{M}_a^{L R}&=&  16 \sqrt{\frac{2}{3}} \pi m_{B_c}^4 C_F  \int_0^1 d x_B d z d x_3 \int_0^{\infty} b_1 b_3 d b_1 d b_3 \phi_B\left(x_B, b_1\right) \phi_D\left(x_3, b_3\right) \phi_0 \non
      &&\times\left\{[r[(1+\eta)(x_B-r_b-1)+\bar\eta x_3+\eta z]\phi^0+\bar\eta\sqrt{\eta(1-r^2)}(1+r_b-x_B-z)(\phi^s+\phi^t)]\right.\non
      &&\left.\times\alpha_s (t_e)h\left(\beta_e, \alpha_a, b_1, b_3\right)  \exp \left[-S_{ef}\left(t_e\right)\right] \right.\non
      &&\left. +r[(1+\eta)(x_B+r_c)-\bar\eta x_3-\eta z]\phi^0-\bar\eta\sqrt{\eta(1-r^2)}(r_c+x_B-z)(\phi^s+\phi^t)\right.\non
      &&\left.\times\alpha_s (t_f) h\left(\beta_f, \alpha_a, b_1, b_3\right)  \exp \left[-S_{ef}\left(t_f\right)\right]\right\},
 \en
The amplitudes from the factorizable annihilation diagrams Fig. 1(g) and 1(h) are listed as
\be
\mathcal{F}_a^{L L}&=&-8 \pi f_{B_c} m_{B_c}^4 C_F  \int_0^1  d z d x_3 \int_0^{\infty} b b_3 d b d b_3  \phi_D\left(x_3, b_3\right)\left\{\left[\bar\eta(\bar\eta x_3+\eta)\phi^0 \right.\right. \non
&& \left.\left.+2r\sqrt{\eta(1-r^2)}(1+\eta+\bar\eta x_3)\phi^s\right]\alpha_s (t_g) h\left(\alpha_a,\beta_g, b, b_3\right) \exp \left[-S_{gh}\left(t_g\right)\right]S_t(x_3) \right.\non
&&\left.+\left[\left[2(1+\eta)rr_c-\bar\eta z+r^2(2x_2\bar\eta-1)\right]\phi^0-\sqrt{\eta(1-r^2)}\left[2rz(\phi^s+\phi^t)\right.\right.\right.\non
&&\left.\left.\left.+(2r-r_c)\bar\eta(\phi^s-\phi^t)\right]\right]\alpha_s (t_h) h\left(\alpha_a, \beta_h, b_3, b\right)  \exp \left[-S_{gh}\left(t_h\right)\right]S_t(z)\right\} ,\\
\mathcal{F}_a^{SP}&= & 16 \pi f_{B_c} m_{B_c}^4 C_F  \int_0^1  d z d x_3 \int_0^{\infty} b b_3 d b d b_3  \phi_D\left(x_3, b_3\right)  \non
&&\times\left\{[r(x_3\bar\eta+2\eta)\phi^0+2\bar\eta\sqrt{\eta(1-r^2)}\phi^s]\alpha_s (t_g)  h_e(\alpha_a, \beta_g, b, b_3)  \exp \left[-S_{gh}\left(t_g\right)\right]S_t(x_3) \right.\non
      && \left.+[[2rz\eta-\bar\eta(r_c-2r)]\phi^0+\sqrt{\eta(1-r^2)}[\bar\eta z(\phi^s-\phi^t)-4rr_c\phi^s]]\right.\non
      &&\left.\times\alpha_s (t_h) h_e\left(\alpha_a, \beta_h, b_3, b\right)  \exp \left[-S_{gh}\left(t_h\right)\right]S_t(z)\right\}.
 \en

By combining the amplitudes from the different Feynman diagrams, the total decay amplitudes for the quasi-two-body decays $B_c \rightarrow  K^{*} D_{(s)} \rightarrow K \pi D_{(s)}$ are given as
\be
\label{d0kstz}
\mathcal{A}\left(B_c^+ \to K^{* +} D^0\to K^0\pi^+ D^0 \right)&= & \ V_{u s} V_{u b}^*\left[a_1 \mathcal{F}_e^{L L}+C_1 \mathcal{M}_e^{L L}\right]+\ V_{c s} V_{c b}^*\left[a_1 \mathcal{F}_a^{L L}+C_1 \mathcal{M}_a^{L L}\right] \non
&& -\ V_{t s} V_{t b}^*\left[\left(C_3+C_9\right)\left(\mathcal{M}_e^{L L}+\mathcal{M}_a^{L L}\right)+\left(C_5+C_7\right)\left(\mathcal{M}_e^{L R}+\mathcal{M}_a^{L R}\right)\right. \non
&& +\left(C_4+\frac{1}{3} C_3+C_{10}+\frac{1}{3} C_9\right)\left(\mathcal{F}_a^{L L}+\mathcal{F}_e^{L L}\right) \non
&& \left.+\left(C_6+\frac{1}{3} C_5+C_8+\frac{1}{3} C_7\right)\left(\mathcal{F}_a^{S P}+\mathcal{F}_e^{S P}\right)\right],\\
\mathcal{A}\left(B_c^+ \to K^{*0} D^{+}\to K^+\pi^-D^+ \right)&=& V_{c s} V_{c b}^*\left[a_1 \mathcal{F}_a^{L L}+C_1 \mathcal{M}_a^{L L}\right]-V_{t s} V_{t b}^*\left[\left(C_3-\frac{1}{2} C_9\right) \mathcal{M}_e^{L L}\right. \non
&& +\left(C_3+C_9\right) \mathcal{M}_a^{L L}+\left(C_5-\frac{1}{2} C_7\right) \mathcal{M}_e^{L R}+\left(C_5+C_7\right) \mathcal{M}_a^{L R} \non
&& +\left(C_4+\frac{1}{3} C_3+C_{10}+\frac{1}{3} C_9\right) \mathcal{F}_a^{L L} \non
&&+\left(C_4+\frac{1}{3} C_3-\frac{1}{2} C_{10}-\frac{1}{6} C_9\right) \mathcal{F}_e^{L L}+\left(C_6+\frac{1}{3} C_5-\frac{1}{2} C_8\right. \non
&& \left.\left.-\frac{1}{6} C_7\right) \mathcal{F}_e^{S P}+\left(C_6+\frac{1}{3} C_5+C_8+\frac{1}{3} C_7\right) \mathcal{F}_a^{S P}\right],\\
\mathcal{A}\left(B_c^+ \to \bar{K}^{*0} D_s^{+}\to K^-\pi^+D_s^{+} \right)&= &   V_{c d} V_{c b}^*\left[a_1 \mathcal{F}_a^{L L}+C_1 \mathcal{M}_a^{L L}\right]- V_{t d} V_{t b}^*\left[\left(C_3-\frac{1}{2} C_9\right) \mathcal{M}_e^{L L}\right. \non
&& +\left(C_3+C_9\right) \mathcal{M}_a^{L L}+\left(C_5-\frac{1}{2} C_7\right) \mathcal{M}_e^{L R}+\left(C_5+C_7\right) \mathcal{M}_a^{L R} \non
&& +\left(C_4+\frac{1}{3} C_3+C_{10}+\frac{1}{3} C_9\right) \mathcal{F}_a^{L L} \non
&& +\left(C_4+\frac{1}{3} C_3-\frac{1}{2} C_{10}-\frac{1}{6} C_9\right) \mathcal{F}_e^{L L}+\left(C_6+\frac{1}{3} C_5-\frac{1}{2} C_8\right. \non
&& \left.\left.-\frac{1}{6} C_7\right) \mathcal{F}_e^{S P}+\left(C_6+\frac{1}{3} C_5+C_8+\frac{1}{3} C_7\right) \mathcal{F}_a^{S P}\right].
\en
Similarly, we can also obtain the total amplitudes for the pure annihilation decays $B_c \rightarrow \ K^{*} h \rightarrow K \pi h $ with $h=K, \pi, \eta^{(\prime)}$ as follows
\be
\mathcal{A}\left(B_c^+ \to \bar{K}^{* 0} K^{+}\to \bar K^0\pi^0 K^+\right)&=& V_{u d}V_{c b}^*\left[a_1 \mathcal{F}_a^{L L}+C_1 \mathcal{M}_a^{L L}\right],\label{kstkpi}\\
\mathcal{A}\left(B_c^+ \to K^{*+} \bar{K}^{0}\to K^0\pi^+\bar{K}^{0} \right)&=&V_{u d}V_{c b}^* \left[a_1 \mathcal{F}_a^{L L}+C_1 \mathcal{M}_a^{L L}\right],\\
\mathcal{A}\left(B_c^+ \to K^{*+} \pi^0\to K^0\pi^+\pi^0\right)&=&V_{u s}V_{c b}^* \left[a_1 \mathcal{F}_a^{L L}+C_1 \mathcal{M}_a^{L L}\right],\\
\mathcal{A}\left(B_c^+ \to K^{*0} \pi^{+}\to K^0\pi^0\pi^+\right)&=&\mathcal{A}\left(B_c^+ \to K^{*+} \pi^0\to K^0\pi^+\pi^0\right),
\en
\be
\mathcal{A}\left(B_c^+ \to K^{*+} \eta\to K^0\pi^+\eta \right)&=&V_{u s}V_{c b}^* \left[a_1 (\mathcal{F}_a^{L L,\eta_q}\cos \phi -\mathcal{F}_a^{L L,\eta_s}\sin \phi)+C_1 (\mathcal{M}_a^{L L,\eta_q}\cos \phi+\mathcal{F}_a^{L L,\eta_s}\sin \phi)\right],\quad\;\;\\
\mathcal{A}\left(B_c^+ \to K^{*+} \eta'\to K^0\pi^+\eta' \right)&=&V_{u s}V_{c b}^* \left[a_1 (\mathcal{F}_a^{L L,\eta_q}\sin \phi +\mathcal{F}_a^{L L,\eta_s}\cos \phi)+C_1 (\mathcal{M}_a^{L L,\eta_q}\sin \phi+\mathcal{F}_a^{L L,\eta_s}\cos \phi)\right],\quad\;\;
\label{ksteta}
\en
where the combinations of the Wilson coefficients $a_1=C_2+C_1 / 3$ and $a_2=C_1+C_2 / 3$, the subscripts $\eta_{q,s}$ represent the two flavor states composing to the physical states $\eta$ and $\eta^{\prime}$ as following
\be
\left(\begin{array}{c}
\eta \\
\eta^{\prime}
\end{array}\right) & = \left(\begin{array}{cc}
\cos \phi & -\sin \phi \\
\sin \phi & \cos \phi
\end{array}\right)\left(\begin{array}{l}
\eta_{q} \\
\eta_{s}
\end{array}\right)
\en
with $\phi=39.3^{\circ}\pm1.0^\circ$ \cite{feld}.

Then the differential decay rate can be described as
\begin{equation}
    \frac{d \mathcal{B}}{d \omega^2}=\tau_{B_c} \frac{\left|\vec{p}_1\right|\left|\vec{p}_3\right|}{64 \pi^3 m_B^3}\left|\mathcal{A}\right|^2,
    \label{bran}
\end{equation}
where $\tau_{B_c}$ is the mean lifetime of $B_{c}$ meson, the kinematic variables
$\left|\vec{p}_1\right|$ and $\left|\vec{p}_3\right|$ denote the magnitudes of the $K$ and the bachelor meson $h$ momenta in the center-of-mass frame of the $K \pi$ pair,
\begin{equation}
\begin{aligned}
\left|\vec{p}_1\right| & =\frac{1}{2} \sqrt{\left[\left(m_K^2-m_\pi^2\right)^2-2\left(m_K^2+m_\pi^2\right) w^2+w^4\right] / w^2}, \\
\left|\vec{p}_3\right| & =\frac{1}{2} \sqrt{\left[\left(m_{B_c}^2-m_h^2\right)^2-2\left(m_{B_c}^2+m_h^2\right) w^2+w^4\right] / w^2}.
\end{aligned}
\end{equation}

{\centering\section{NUMERICAL RESULTS }}

The adopted input parameters in our numerical calculations are summarized as following (the QCD scale, the masses, the decay constants and the widths are in units of $\mathrm{GeV}$, the $B_c$ meson lifetime is in units of ps) \cite{pdg}
\begin{equation}
\begin{aligned}
&\Lambda_{Q C D} =0.25, m_{B_c^+}=6.274, m_b =4.8,m_{K^{\pm}}=0.494, m_{K^0}=0.498,\\
&m_{\pi^{\pm}}=0.140, m_{\pi^0} =0.135, m_{K^{* 0}}=0.89555, m_{K^{*\pm}}=0.89176, \\
&f_{B_c}=0.489, f_{K^*}=0.217, \Gamma_{K^{* 0}}=0.0462, \Gamma_{K^{*\pm}}=0.0514, \tau_{B_c}=0.51. \\
\end{aligned}
\end{equation}
As to the Cabibbo-Kobayashi-Maskawa (CKM) matrix elements, we employ the Wolfenstein parametrization with the inputs\cite{pdg}
\begin{equation}
\begin{aligned}
& \lambda=0.22500\pm0.00067, \quad A=0.826^{+0.018}_{-0.015}, \\
& \bar{\rho}=0.159\pm0.010, \quad \bar{\eta}=0.348\pm0.010.
\end{aligned}
\end{equation}
By using the differential branching ratio in Eq.(\ref{bran}) and the squared amplitudes in Eqs.(\ref{d0kstz})-(\ref{ksteta}), integrating over the
full $K \pi$ invariant mass region $\left(m_K+m_\pi\right) \leq \omega \leq (M_{B_{c}}-m_h)$ with $h= D_{(s)}, K, \pi, \eta^{(\prime)}$,
we obtain the branching ratios for these quasi-two-body decays as
\be
Br\left(B_c^+ \to \ K^{*+} D^{0}\rightarrow\ K^{0}\pi^+ D^{0}\right) & =  & \left(8.74_{-1.43-0.03-0.01-1.03}^{+1.30+0.04+0.00+1.61} \right)\times 10^{-7}, \label{bctokzd}\\
Br\left(B_c^+ \to \ K^{*0} D^{+} \rightarrow \ K^+ \pi^- D^+\right) & = & \left( 14.0_{-3.06-0.14-0.14-2.35}^{+0.73+0.02+0.00+2.20}  \right)\times 10^{-7}, \\
Br\left(B_c^+ \to \bar K^{*0} D_s^{+} \rightarrow K^{-}\pi^+ D_s^{+}\right) & = & \left( 1.17_{-0.22-0.00-0.00-0.44}^{+0.19+0.03+0.05+0.00}  \right)\times 10^{-7}, \\
Br\left(B_c^+ \to \bar K^{*0}\ K^{+}\rightarrow \bar K^{0}\pi^0\ K^{+}\right) & = & \left( 3.32_{-0.00-0.13-0.00-0.00}^{+0.01+0.14+0.01+1.37}  \right)\times 10^{-7}, \\
Br\left(B_c^+ \rightarrow \ K^{*+}\bar{K^{0}}\rightarrow\ K^{0}\pi^+\bar{K^{0}}\right) & = & \left( 1.30 _{-0.01-0.09-0.06-0.37}^{+0.15+0.05+0.15+0.00}  \right)\times 10^{-7}, \\
Br\left(B_c^+ \rightarrow \ K^{*0} \pi^{+} \rightarrow \ K^0 \pi^0\pi^{+}\right) & = & \left( 0.74_{-0.00-0.01-0.00-0.24}^{+0.00+0.01+0.00+1.09}  \right)\times 10^{-8}, \\
Br\left(B_c^+ \rightarrow \ K^{*+} \pi^{0} \rightarrow \ K^{0}\pi^+\pi^{0}\right) & =  & \left(0.74_{-0.01-0.01-0.01-0.41}^{+0.00+0.01+0.01+0.81}  \right)\times 10^{-8}, \\
Br\left(B_c^+ \rightarrow \ K^{*+} \eta \rightarrow \ K^{0}\pi^+\eta\right) & = & \left( 0.50_{-0.00-0.03-0.08-0.13}^{+0.01+0.03+0.09+0.35}  \right)\times 10^{-8}, \\
Br\left(B_c^+ \rightarrow \ K^{*+} \eta' \rightarrow \ K^{0}\pi^+\eta' \right) & = &  \left(1.58_{-0.01-0.02-0.09-1.31}^{+0.00+0.00+0.04+0.00}  \right)\times 10^{-8},
\label{bcksteta}
\en
where the first error is from the $B_c$ meson shape parameter uncertainty $\omega_{B_c}=1.0\pm0.1$ GeV, the following two errors come
from the Gegenbauer coefficients in the $K \pi$ pair distribution amplitudes
$ a_{1 K^*}^{\parallel}=0.05 \pm 0.02, a_{2 K^*}^{\parallel}=0.15 \pm 0.05$, and the last one is induced by varying the hard scale $t$ from $0.75t$ to $1.25t$ (without changing $1/b_i$) and
QCD scale $\Lambda_{QCD}=0.25 \pm 0.05$ GeV, which characterizes the next-to-leading-order effect in the PQCD approach.
One can find that the errors induced by $a_{1 K^*}^{\parallel}, a_{2 K^*}^{\parallel}$ are from a few percent to $15\%$ for the most of these
considered decays. For the pure annihilation decay modes, the error stemming from the uncertainty of the $B_c$ meson shape
parameter $\omega_{B_c}$ is small. It can be roughly understood from the analytical formulas Eqs.(\ref{kstkpi})-(\ref{ksteta}).
Because the Wilson coefficient $a_1=C_2+C_1/3$ is larger than $C_1$, the main contribution comes from the factorizable annihilation amplitude
$\mathcal{F}^{LL}_a$, where the terms about $\omega_{B_c}$ are not involved.
The situation is a little different with the decay $B_c^+ \to \ K^{*+} D^{0}\to\ K^{0}\pi^+ D^{0}$, where the error induced by $\omega_{B_c}$
can reach $16\%$.
 By comparison, the branching ratios of these pure annihilation
decays are more sensitive to
the variation of the hard scale $t$ and the QCD scale $\Lambda_{Q C D}$. It means that these decays might be sensitive to the higher order corrections.
The errors arise from the uncertainties of the parameters, for
instance, the Wolfenstein parameters, the pole mass of $m_{K^*}$ and width $\Gamma_{K^*}$, are very small and have been neglected.

If we assume the isospin conservation for the strong decays $K^*\to K\pi$, namely
\begin{equation}
\begin{gathered}
\frac{\Gamma\left(K^{* 0} \rightarrow K^{+} \pi^{-}\right)}{\Gamma\left(K^{* 0} \rightarrow K \pi\right)}=2 / 3, \frac{\Gamma\left(K^{* 0} \rightarrow K^0 \pi^0\right)}{\Gamma\left(K^{* 0} \rightarrow K \pi\right)}=1 / 3, \\
\frac{\Gamma\left(K^{*+} \rightarrow K^0 \pi^{+}\right)}{\Gamma\left(K^{*+} \rightarrow K \pi\right)}=2 / 3, \frac{\Gamma\left(K^{*+} \rightarrow K^{+} \pi^0\right)}{\Gamma\left(K^{*+} \rightarrow K \pi\right)}=1 / 3 .
\end{gathered}
\end{equation}
Under the narrow width approximation relation, the branching ratio of each quasi-two-body decay can be related with that of
the corresponding two-body decay using a simple formula. Take the decay $B_c^+ \to \ K^{*+} D^{0}\rightarrow\ K^{0}\pi^+ D^{0}$ as an example, the formula
can be expressed as
\be
Br\left(B_c^+ \rightarrow \ K^{*+} D^{0} \rightarrow \ K^{0}\pi^+ D^{0} \right)
= Br\left(B_c^+ \rightarrow  \ K^{*+} D^{0}\right) \cdot Br\left(K^{*+} \rightarrow K^{0} \pi^{+}\right).
\en
\begin{table}[h]
\caption{The CP averaged branching ratios for the two-body decays $B_c^+ \to  K^{*}h$ with $h= D_{(s)}, K, \pi, \eta^{(\prime)}$ . The errors are the same with those given in Eqs.(\ref{bctokzd})-(\ref{bcksteta}).}
    \centering
    \renewcommand\arraystretch{2}
    \begin{tabular}{cccc}
    \hline
          Decay modes  &  This Work  & Two-body framework \cite{Rui:2011qc,Liu:2009qa}  &  RQCM \cite{Liu:1997hr}   \\ \hline

           $B_c^{+} \to K^{*+} D^0$  &  $\left(1.31  _{-0.21-0.01-0.00-0.16}^{+0.20+0.01+0.00+0.24}\right) \times 10^{-6}$
             &  $ (2.59_{-0.30-0.08-0.08}^{+0.27+0.09+0.15}) \times 10^{-6} $
             &  $3.47\times 10^{-6}$    \\ \hline
           $B_c^{+} \rightarrow K^{* 0} D^{+}$  &   $\left(2.10 _{-0.46-0.02-0.02-0.35}^{+0.11+0.00+0.00+0.33}\right) \times 10^{-6}$
             &  $ (1.91_{-0.25-0.00-0.07}^{+0.33+0.01+0.07}) \times 10^{-6} $
             &  $2.88\times10^{-6}$   \\ \hline
           $B_c^{+} \rightarrow \bar{K}^{* 0} D_s^{+}$  &  $(1.76_{-0.33-0.00-0.00-0.66}^{+0.28+0.04+0.07+0.00}) \times 10^{-7} $
             &  $ (1.4_{-0.2-0.1-0.1}^{+0.2+0.0+0.1}) \times 10^{-7} $
             &  $1.0\times10^{-7}$    \\ \hline
 $ B_c^{+} \rightarrow \bar{K}^{* 0} K^{+} $  &  $(9.97_{-0.00-0.39-0.00-0.00}^{+0.03+0.42+0.04+4.10}) \times 10^{-7} $
             &  $ (10.0_{-0.6-3.3-0.2}^{+0.5+1.7+0.0}) \times 10^{-7} $
             & -   \\ \hline
$ B_c^{+} \rightarrow K^{*+} \bar{K}^0 $  &  $\left(1.95 _{-0.01-0.14-0.09-0.55}^{+0.21+0.07+0.22+0.00}\right) \times 10^{-7} $
             &  $ (1.8_{-0.1-2.1-0.0}^{+0.7+4.1+0.1}) \times 10^{-7} $
             &  -    \\ \hline
$ B_c^{+} \rightarrow K^{* 0} \pi^{+} $  &  $\left(2.23 _{-0.00-0.02-0.01-0.73}^{+0.00+0.03+0.00+3.27}\right) \times 10^{-8}$
             &  $ (3.3_{-0.2-0.4-0.1}^{+0.7+0.4+0.2}) \times 10^{-8} $&- \\ \hline
$ B_c^{+} \rightarrow K^{*+} \pi^0 $  &  $\left(1.11 _{-0.01-0.02-0.02-0.62}^{+0.00+0.01+0.02+1.21}\right) \times 10^{-8} $
             &  $ (1.6_{-0.1-0.1-0.0}^{+0.4+0.3+0.1}) \times 10^{-8} $ \\ \hline
        $ B_c^{+} \rightarrow K^{*+} \eta $  &  $\left(0.75 _{-0.00-0.04-0.12-0.20}^{+0.01+0.04+0.13+0.52}\right) \times 10^{-8} $
             &  $ (0.9_{-0.0-0.2-0.0}^{+0.1+0.6+0.0}) \times 10^{-8} $ &-\\ \hline
           $ B_c^{+} \rightarrow K^{*+} \eta^{\prime} $  &  $\left(2.37_{-0.02-0.03-0.14-1.97}^{+0.00+0.00+0.06+0.00}\right) \times 10^{-8} $
             &  $ (3.8_{-1.1-0.6-0.0}^{+1.1+1.0+0.0}) \times 10^{-8} $ & -\\ \hline\hline
 \end{tabular}
\label{tabbran}
\end{table}
Then we can obtain the branching ratios of the relevant two-body decays from those of the considered quasi-two-body decays,
which are listed in Table \ref{tabbran}. In these considered decays,  $B_c^+ \to K^{*+}D^{0}$ and $B_c^+ \to K^{*0}D^{+}$  have the largest branching ratios and reach $10^{-6}$,
which are possible measured by the future High-energy LHC (HE-LHC) and
High-luminosity LHC (HL-LHC) experiments \cite{run3}. From our calculations, we find that in these decays with $D_{(s)}$ meson involved,
the penguin amplitudes are dominated. Although the values of the CKM matrix elements $V_{cb}V_{cs(d)}$ and $V_{tb}V_{ts(d)}$ are close to each other, the tree contributions associated with the CKM matrix elements $V_{cb}V_{cs(d)}$ are from the annihilation type amplitudes and very tiny.
For example, such contributions are only about $2.9(1.1)\%$ of the total branching ratio for the decay $B_c^+ \to K^{*0}D^{+}
(B_c^+ \to \bar K^{*0}D^{+}_s)$. There are
some differences in the decay $B_c^+ \to K^{*+}D^{0}$, which receives two kinds of tree contributions: one is associated with the CKM matrix
elements $V^*_{ub}V_{us}$, the other is associated with the CKM elements $V^*_{cb}V_{cs}$. Although $V^*_{ub}V_{us}$ is smaller than $V^*_{cb}V_{cs}
(\frac{V^*_{ub}V_{us}}{V^*_{cb}V_{cs}}=0.0215)$, the former is involved in the (non-)factorizable emission amplitude
$\mathcal{F}_e^{L L}(\mathcal{M}_e^{L L})$
and the latter is involved in the (non-)factorizable annihilation amplitude $\mathcal{F}_a^{L L}(\mathcal{M}_a^{L L})$ shown in Eq.(\ref{d0kstz}).
It is interesting that the differences from the amplitudes are huge enough to compensate the differences from the CKM elements. In fact, the
proportions of these three kinds of contributions (coming from $V^*_{ub}V_{us}, V^*_{cb}V_{cs}$ and $V^*_{tb}V_{ts}$ ) in the total branching
ratios are $1:0.24:0.06$. So the tree contributions coming from CKM matrix elements $V^*_{ub}V_{us}$ are more important than those from $V^*_{cb}V_{cs}$
in the decay $B_c^+ \to K^{*+}D^{0}$.
Although these tree contributions are not much helpful to increase the branching ratio, they are important to the direct CP violation of the decay $B_c^+ \to K^{*+}D^{0}$.
Since the absence of the tree contribution from the factorizable and non-factorizable emission amplitudes in the decays $B_c^+ \to K^{*0}D^{+}$ and $B_c^+ \to \bar K^{*0}D^{+}_s$, the direct CP
asymmetries in these two decays may be smaller. We will make a detailed discussion about this topic in the latter.
As to the decay $B_c^+ \to \bar K^{*0}D^{+}_s$, its analytical formulas of the
 tree (penguin) amplitudes
are almost same with those of the decay $B_c^+ \rightarrow K^{*0}D^{+}$ (the differences are from the wave functions of $D$ and $D_s$),
while the value of the corresponding CKM elements $V_{cd}
(V_{td})$ is only about 0.2 times of that of $V_{cs} (V_{ts})$.
So the branching ratio of the decay $B_c^+ \to \bar K^{*0}D^{+}_s$ is much smaller and only at $10^{-7}$ order.
Compared with the branching ratios obtained in the previous PQCD calculations under the two-body framework \cite{Liu:2009qa,Rui:2011qc},
one can find that three-body and two-body calculations about these decays are consistent with each other, it supports
the PQCD approach to exclusive hadronic $B_c$ meson decays.  For the decay $B_c^+ \rightarrow K^{*+}D^{0}$,
its branching ratio is smaller than the result given by
the relativistic constituent quark model (RCQM) \cite{Liu:1997hr}, but is much larger than $0.68\times10^{-7}$ predicted by the light front
quark model (LFQM) \cite{choi} and
 $1.36\times10^{-7}$ given by the Salpeter method \cite{hffu}. For the contributions from annihilation diagrams and  penguin diagrams are missed
in the LFQM and Salpter method, so the decays dominated by the annihilation and penguin contributions might not be well predicted
by these approaches. It is not surprise that $Br(B_c^+ \to K^{*0}D^{+})=1.59\times10^{-7}$ and
$Br(B_c^{+} \to \bar{K}^{* 0} D_s^{+})=2.09\times10^{-8}$ given by the Salpeter method are almost one order of magnitude smaller than the PQCD predictions.
It is meaningful to clarify these controversies in the future LHCb experiments.
For the pure annihilation decays $B_c\to K^*h$ with $h$ representing a light pseudoscalar meson $K, \pi$ or $ \eta^{(\prime)}$,
there are two decay modes, one is strange decay ($\Delta S=1$) corresponding to the smaller CKM
matrix element
$V_{us}\sim 0.22$, which refers to $B_c^+ \to K^{*0}\pi^{+}, K^{*+}\pi^{0}, K^{*+}\eta^{(\prime)}$, the other is non-strange
decay ($\Delta S=0$) corresponding to the larger CKM matrix element $V_{us}\sim 1$, which refers to $B_c^+ \to \bar K^{*0} K^{+}, K^{*+} \bar K^{0}$.
One can find that the branching ratios for the $\Delta S=0$ channels  are much larger than those for the $\Delta S=1$ decays. For these two
$\Delta S=0$ processes, the decay $B_c^+ \to \bar K^{*0} K^{+}$ has the larger branching ratio, which is near $10^{-6}$ and is possible observed by the future LHCb experiments. It is interesting that this result is consistent with the estimation from the SU(3) flavour symmetry \cite{genon}. Although both of the decays $B_c^+ \to \bar K^{*0} K^{+}$ and $B_c^+ \to  K^{*+} \bar K^{0}$ belong to the same decay mode,
there exists large gap between their branching ratios.
While it is very different for the case of $Br(B_u^+ \to \bar K^{*0} K^{+})$ and $Br(B_u^+ \to  K^{*+} \bar K^{0})$, which are close to each other
 predicted by many theoretical approaches, such as the PQCD approach \cite{chai}, the QCD factorization approach (QCDF) \cite{chua} and the soft-collinear effective theory (SCET) \cite{wangw}. Such abnormality shows
significant difference for the annihilation amplitudes between the $B$ (heavy-light system) and $B_c$ (heavy-heavy system) decays.
If this point can be clarified by the experiments, it is helpful to further improve our understanding of the annihilation contributions.

\begin{figure}[h]
\centering
    \begin{minipage}[t]{0.32\textwidth}
        \centering
        \includegraphics[width=1\textwidth]{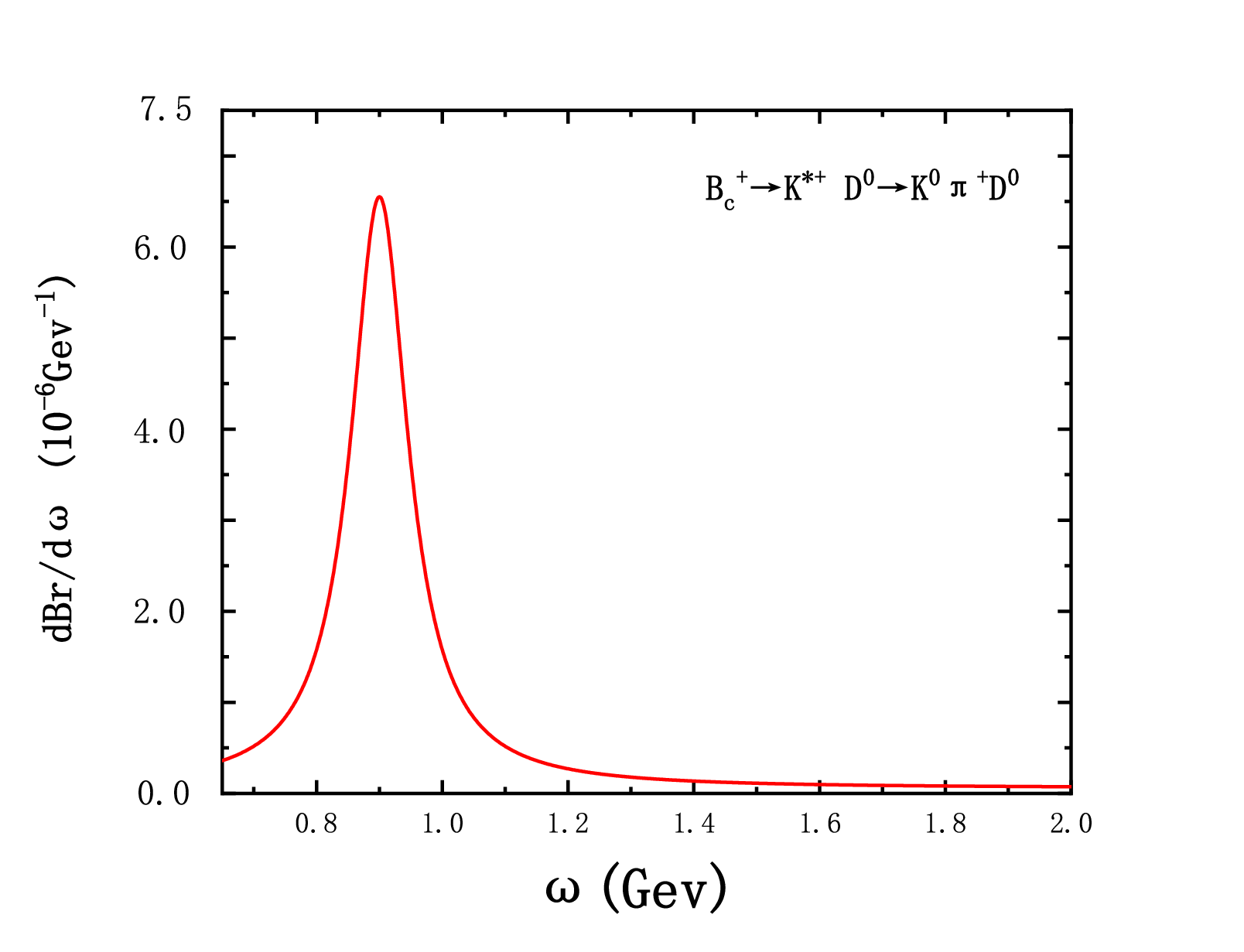}
        \centerline{(a)}
    \end{minipage}
    \begin{minipage}[t]{0.32\textwidth}
         \centering
         \includegraphics[width=1\textwidth]{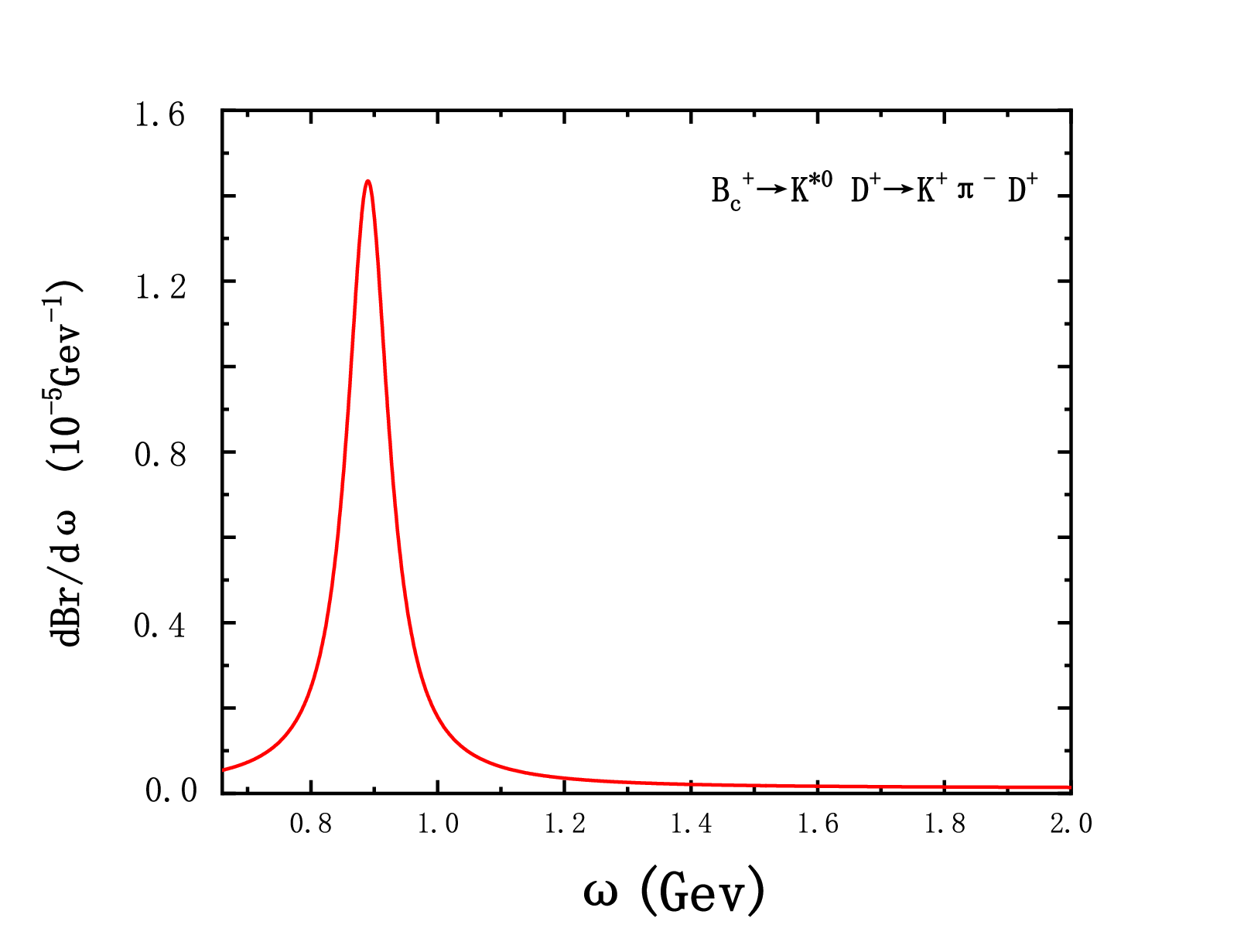}
         \centerline{(b)}
    \end{minipage}
    \begin{minipage}[t]{0.32\textwidth}
         \centering
         \includegraphics[width=1\textwidth]{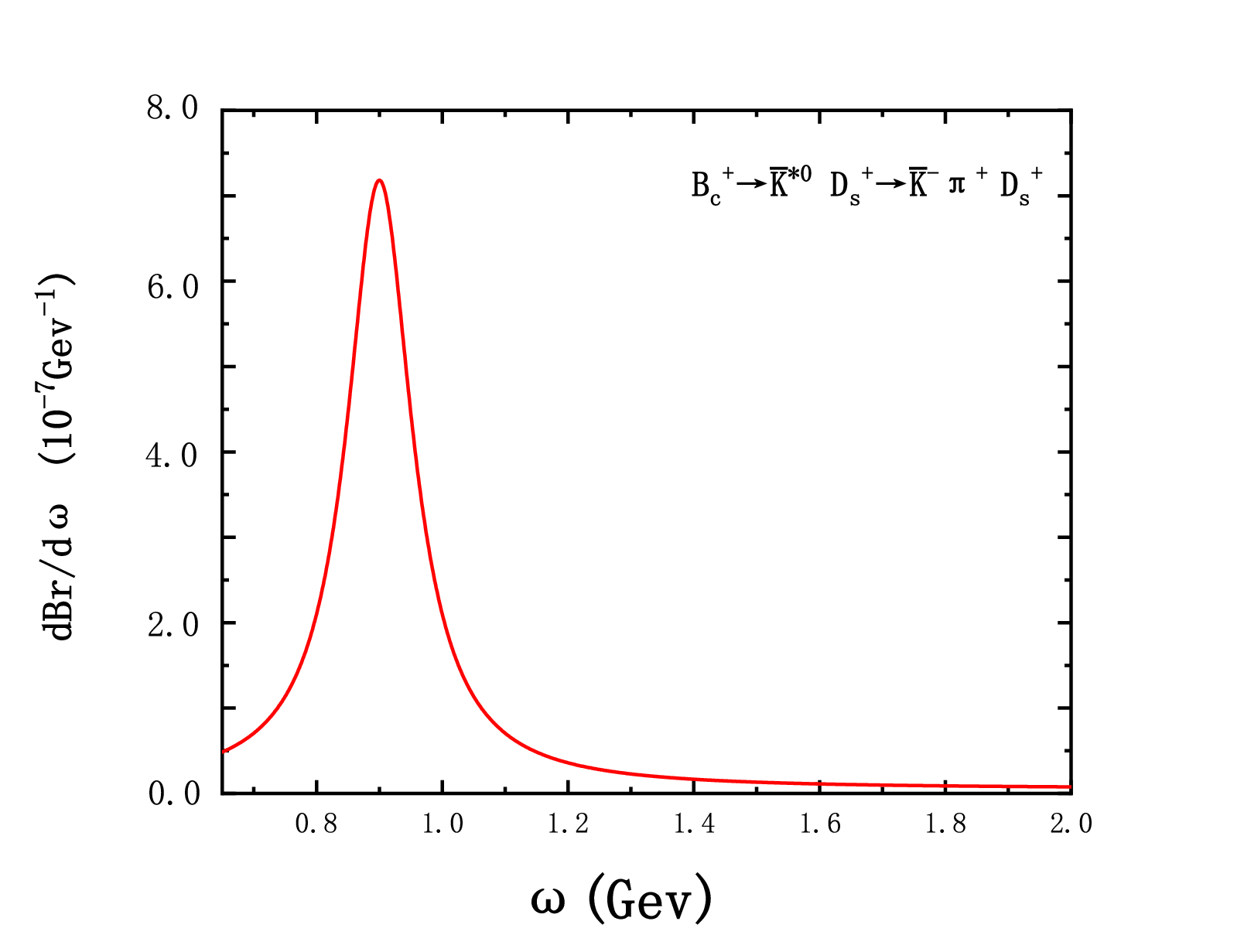}
         \centerline{(c)}
    \end{minipage}
    \caption{ The predicted $B_c\to K^* D_{(s)}\to K \pi D_{(s)}$ decay spectra in the
$K\pi$ invariant mass.}
        \label{diffbranch}
   \end{figure}
The decay amplitudes of the quasi-two-body decays depend on the $K\pi$ invariant mass, which are different from the fixed kinematics in
the two-body decays. So we can plot the differential distribution of the branching ratios, shown in Fig. \ref{diffbranch}, where we take
the decays with $D_{(s)}$ involved as examples. One can see that
the differential branching ratios for these three decays  exhibit peaks at the $K^*$ meson mass. The main portion of the branching ratios lies
in the region around the pole mass of the $K^*$ resonance as expected. For example, the branching ratio
obtained by integrating over $\omega$ in the range $m_{K^*}-\Gamma_{K^*}$ to $m_{K^*}+\Gamma_{K^*}$ is about $70\%$ of the total decay rate for
the channel $B_c^+ \to \ K^{*+} D^{0}\to\ K^{0}\pi^+ D^{0}$. Although
 we plot the differential branching ratios versus the invariant mass $\omega$ in the range $m_K+m_\pi$ to 2GeV, the  contributions from
 the energy region $\omega>1.2$ GeV can be neglected safely.

Now we turn to the evaluations of the CP violation for the decays with $D_{(s)}$ meson involved,
the direct CP violations induced from the interference between the tree and penguin amplitudes, can be defined as
\begin{equation}
    A_{CP}=\frac{\Gamma\left(B_c^- \to \bar f \right)-\Gamma\left(B_c^+ \to f \right)}{\Gamma\left(B_c^- \to \bar f \right)
    +\Gamma\left(B_c^+ \to f \right)},
    \label{cpeq}
\end{equation}
where $\bar f$ is the CP conjugated final state of $f$.  The numerical results for these direct CP asymmetries are given as
\be
A_{CP}\left(B_c^+ \to K^{*+} D^{0} \to  K^{0}\pi^+D^{0}\right) & =&\left(-14.6_{-0.98-0.53-0.14-0.00}^{+2.73+0.43+0.00+8.76}\right) \times 10^{-2}, \\
A_{CP}\left(B_c^+ \to  K^{*0} D^{+} \to K^+ \pi^- D^+\right) & =&\left(-0.139_{-0.00-0.01-0.02-0.00}^{+0.02+0.01+0.02+0.13}\right) \times 10^{-2}, \\
A_{CP}\left(B_c^+ \to \bar K^{*0} D_s^{+} \to K^-\pi^+D_s^{+}\right) & =&\left(1.07_{-0.42-0.24-0.09-7.61}^{+0.22+0.13+0.18+0.00}\right) \times 10^{-2},
\en
where the errors are the same with those given in Eqs.(\ref{bctokzd})-(\ref{bcksteta}). Unlike the branching ratio, the direct
CP asymmetry is not sensitive to the parameters in DAs, but suffer from large uncertainties due to the hard scale $t$ and the
QCD scale $\Lambda_{Q C D}$, which can be reduced by including the high order corrections.
\begin{table}[h]
 \caption{The direct CP violation ($\times 10^{-2}$) of the decays
 $B_c^+ \to K^{*+}D^{0}$ , $B_c^+ \to K^{*0}D^{+}$ and $B_c^+ \to \bar K^{*0}D_s^{+}$, where the various errors have been added in quadrature. By comparison, we also give the results from the PQCD appraoch in the two-body framework \cite{Rui:2011qc}, the Salpeter method \cite{hffu} and the RCQM \cite{Liu:1997hr}.  }
    \label{tab:my_label}
    \centering
    \renewcommand\arraystretch{2}
    \begin{tabular}{ccccc}
    \hline\hline
      Decay modes  &  this work      &   PQCD (two-body framework) \cite{Rui:2011qc} &  Salpeter method \cite{hffu}   &   RCQM \cite{Liu:1997hr}\\ \hline

$ B_c^+ \to K^{*+} D^{0}  $   &  $-14.6_{-1.12}^{+9.19} $ & $-66.2_{-6.5}^{+15.2}$ &   $-25.5$
   &   -6.22 \\
\hline
$ B_c^+ \to  K^{*0} D^{+}   $   &   $-0.139_{-0.02}^{+0.13}$ &    $3.5_{-0.86}^{+0.71}$    &   $-0.53$     &   -0.822\\ \hline
$ B_c^+ \to \bar K^{*0} D_s^{+} $   &   $1.07_{-7.63}^{+0.31}$  &    $61.0_{-14.65}^{+7.91}$    &   $9.04$    &   13.3\\ \hline\hline
 \end{tabular}
 \label{cp}
\end{table}
From the numerical results, we found the following points:
\begin{enumerate}
\item Under the three-body framework calculations, there exist no so large direct CP violations in the the decays
$B_c \to D^0 K^{*+}, D_s^{+} \bar K^{* 0}$, which were predicted as more than $60\%$ in magnitude by the previous PQCD calculations
\cite{Rui:2011qc}. Our predictions are more comparable with those given by the Salpeter method and the RCQM. These results can be clarified by the
future LHCb experiments.
\item
Compared with $B_c^+ \to  K^{*0} D^{+} \to K^+ \pi^- D^+$, the decay $B_c^+ \to K^{*+} D^{0} \to  K^{0}\pi^+D^{0}$ receives more tree amplitudes contributions, which come from not
only the emission diagrams but also the annihilation diagrams. Although the emission factorizable amplitude $F^{LL}_e$
 is suppressed by the CKM matrix elements $V_{us}V^*_{ub}$, it still provides a strong
interference with the penguin amplitudes  because of the large Wilson coefficient $a_1=C_2+C_1/3$. So there exists more significant
direct CP violation
in the decay $B_c^+ \to K^{*+} D^{0} \to  K^{0}\pi^+D^{0}$ than that of the decay $B_c^+ \to  K^{*0} D^{+} \to K^+ \pi^- D^+$ as we expected.
\item
As to the decay $B_c^+ \to  K^{*0} D^{+} \to K^+ \pi^- D^+$, although the CKM matrix element products $V_{cs}V^*_{tb}$ and $V_{ts}V^*_{tb}$
associated with the tree and penguin amplitudes, respectively, are almost equal to each other, the tree contributions from the annihilation type
amplitudes are very small and about two orders lower than the penguin contributions. So the
interference between these two kinds of contributions is weak, which induces much smaller direct CP violation.
\item
The amplitudes of the decay $B_c^+ \to \bar K^{*0} D_s^{+} \to K^-\pi^+D_s^{+} $ can be obtained from those of the channel $B_c^+ \to  K^{*0} D^{+} \to K^+ \pi^- D^+$ by replacing
$D^+ (V_{ts}, V_{cs})$ with $D^+_s (V_{{td}}, V_{cd})$. The total decay amplitudes for these two decays can be rewritten as
\be
\mathcal{A}=V^*_{cb}V_{cq}T-V^*_{tb}V_{tq}P=V^*_{cb}V_{cq}T\left[1+ze^{i(\alpha+\delta)}\right],
\en
where $T$ and $P$ are the tree and penguin amplitudes, $\alpha$ and $\delta$ are the weak and strong phases, respectively.
The parameters $z$ and $\alpha$ are defined as
\be
z=|\frac{V^*_{tb}V_{tq}}{V^*_{cb}V_{cq}}\frac{P}{T}|,\quad \alpha=arg[-\frac{V^*_{tb}V_{tq}}{V^*_{cb}V_{cq}}],
\en
with $q=d(s)$ for the decay $B_c^+ \to \bar K^{*0} D_s^{+} \to K^-\pi^+D_s^{+} (B_c^+ \to  K^{*0} D^{+} \to K^+ \pi^- D^+)$.
Then the direct CP asymmetries are
\be
A_{CP}=\frac{2z\sin{\alpha}\sin\delta}{z^2+1+2\cos{\alpha}\cos{\delta}}.
\en
 As the weak phases are measured as
$arg[-\frac{V^*_{tb}V_{td}}{V^*_{cb}V_{cd}}]\sim-0.40$ and $arg[-\frac{V^*_{tb}V_{ts}}{V^*_{cb}V_{cs}}]\sim0.02$ \cite{pdg}, their corresponding sine values
are about $-0.39$ and $0.02$. So one can find that the the size of $A_{CP}(B_c\to \bar K^{* 0} D_s^{+})$ is larger
than that of $A_{CP}(B^+_c \to K^{*0} D^+)$ because of the larger weak phase shown in Table \ref{cp}. Their opposite signs
 are from the weak phases with opposite signs.
\end{enumerate}
Last, we plot the differential distributions of the direct CP violations for the decays $B^+_c\to K^{*+}D^0 \to K^0\pi^+D^0,
B^+_c \to K^{*0}D^+ \to  K^+\pi^-D^+$ and $B_c\to  \bar K^{* 0}D_s^{+} \to  K^-\pi^+D_s^+$. The measured CP violation
is just a number in the two-body framework, where the $K^*$ resonance mass is fixed to $m_{K^*}$ during the calculations. While the direct
CP violation in the three-body framework is dependent on the $K\pi$ invariant mass $\omega$. So the total direct CP asymmetry is the integration of the $A_{CP}$ differential distribution over $\omega$. The integrated direct CP asymmetry for the quasi-body decays may be very different with that obtained in the two-body framework, that is to say, the latter may be overestimated or underestimated compared with the data. In view of this point, the three-body framework should be more appropriate for studying the quasi-two-body decays.
Here, we also find that the differential distribution curve for $A_{CP}(B_c\to \bar K^{* 0}D_s^{+}\to K^-\pi^+D_s^+)$ lies in the positive value region, which is
contrary to the cases of $A_{CP}(B_c\to K^{*+} D^0  \to K^0\pi^+ D^0 )$ and $A_{CP}(B^+_c \to K^{*0}D^+  \to  K^+\pi^-D^+)$. It is mainly because of
the differences from the weak phases.

\begin{figure}[h]
\centering
    \begin{minipage}[t]{0.32\textwidth}
        \centering
        \includegraphics[width=1\textwidth]{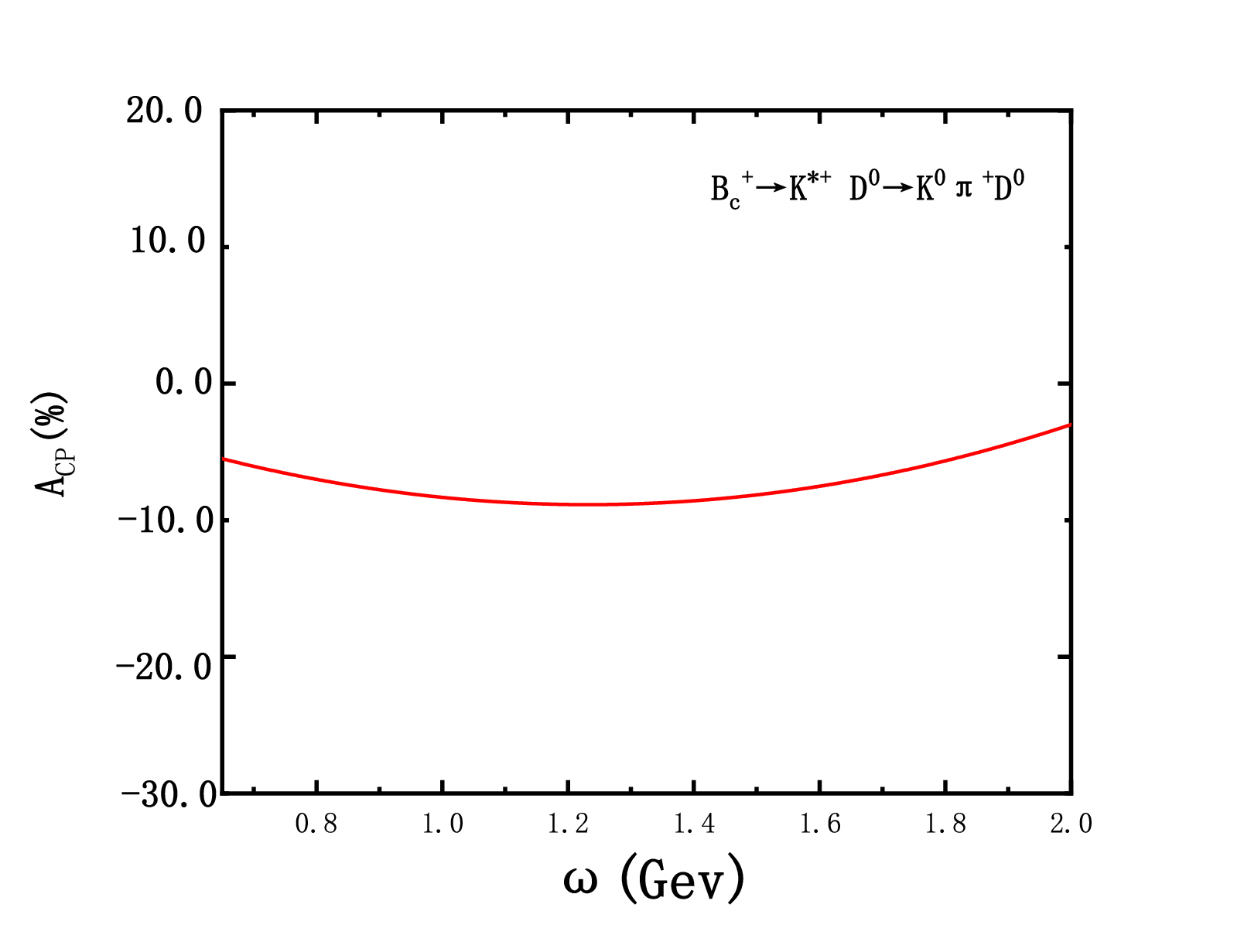}
        \centerline{(a)}
    \end{minipage}
    \begin{minipage}[t]{0.32\textwidth}
         \centering
         \includegraphics[width=1\textwidth]{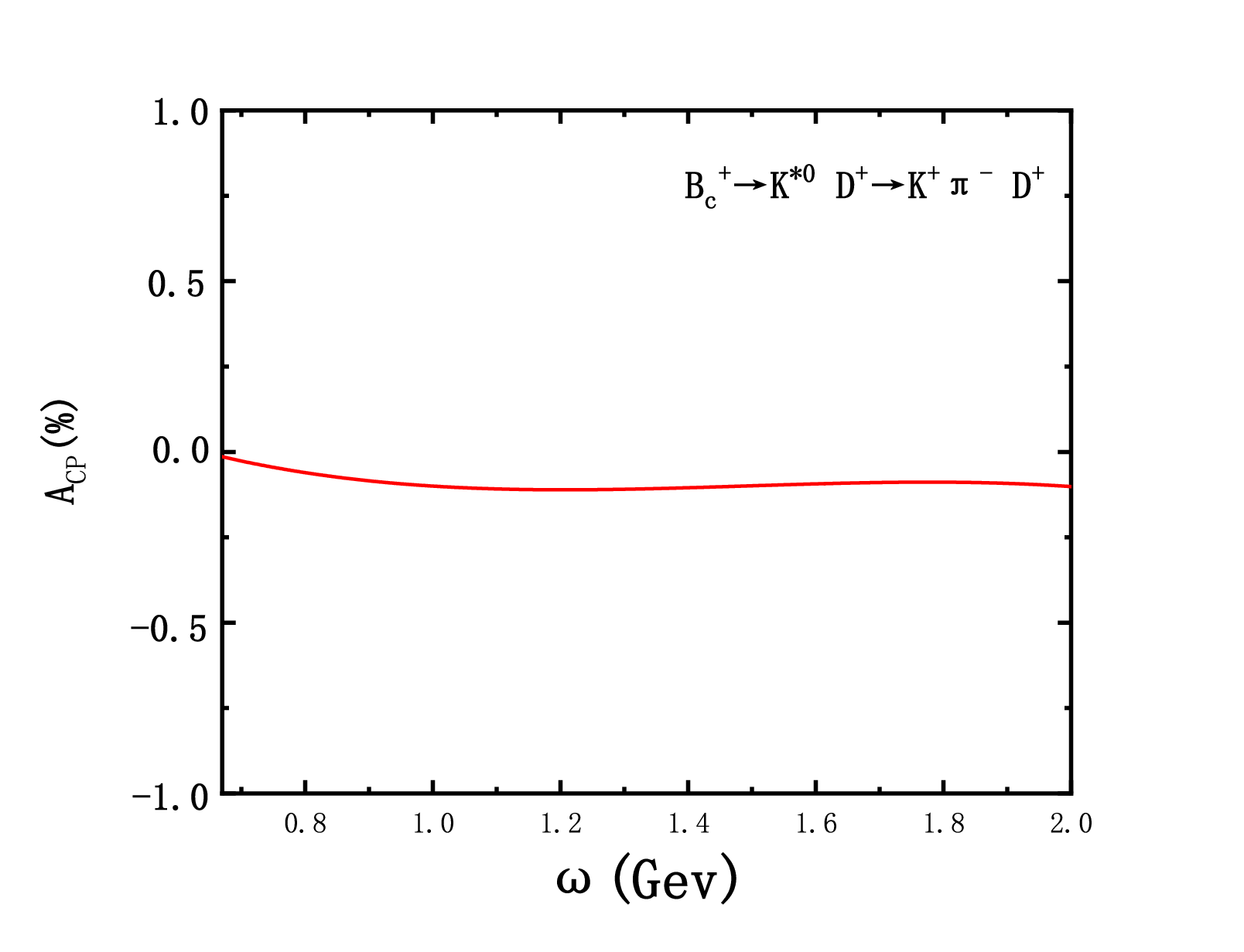}
         \centerline{(b)}
    \end{minipage}
    \begin{minipage}[t]{0.32\textwidth}
         \centering
         \includegraphics[width=1\textwidth]{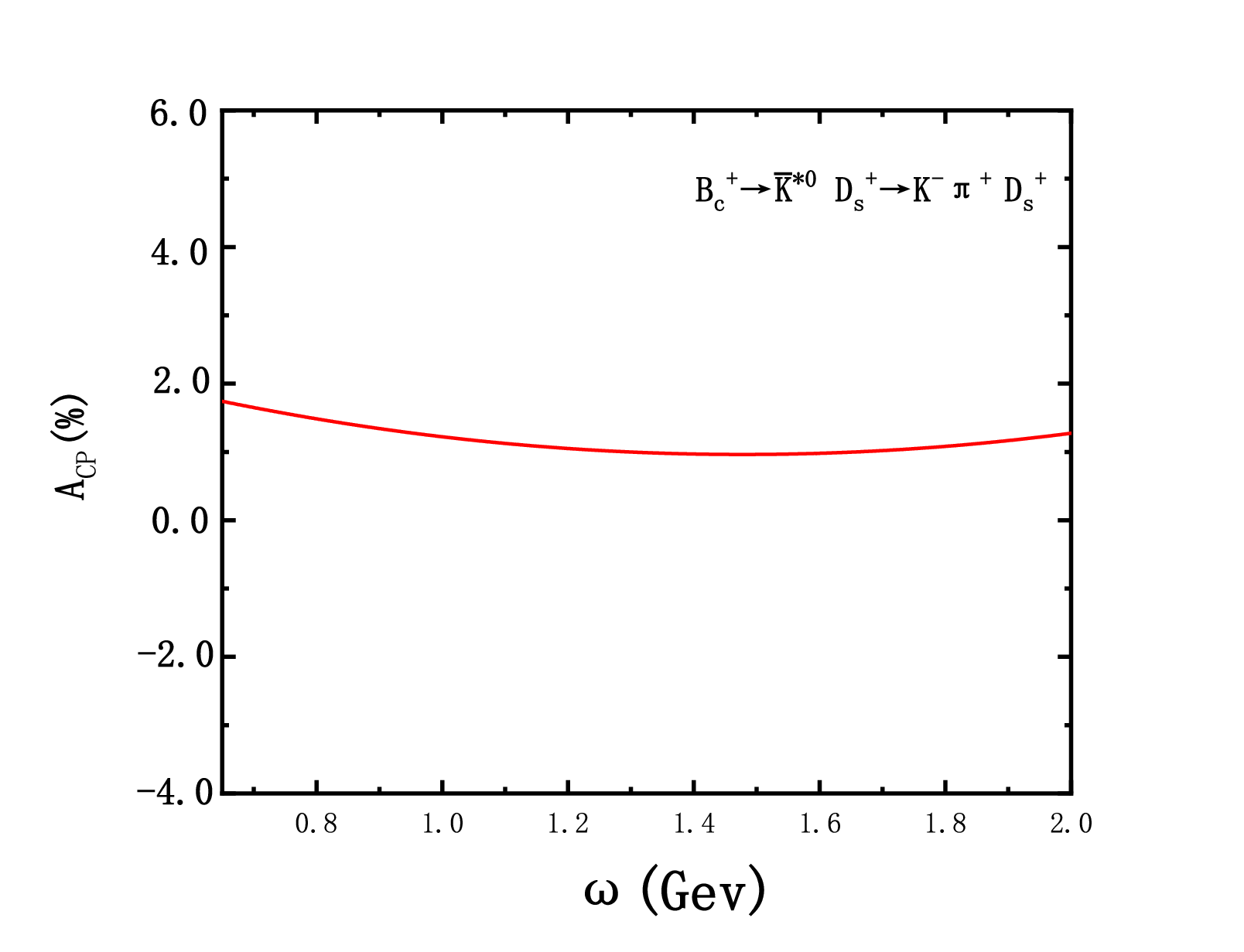}
         \centerline{(c)}
    \end{minipage}
    \caption{ The differential distributions of $A_{CP}$ in $\omega$ for the decays $B^+_c\to K^{*+} D^0  \to K^0\pi^+ D^0 , B^+_c\to K^{*0} D^+  \to K^+\pi^- D^+$ and $B^+_c\to  \bar K^{*0}D^+_s \to K^-\pi^+D^+_s$. }
        \label{}
   \end{figure}

{\centering\section{SUMMARY}}

In this paper, we studied the quasi-two-body decays
$B_c \to \ K^{*} h \to \ K\pi h $ with $h = D, D_s, K, \pi, \eta, \eta'$ by using the PQCD approach. Under the quasi-two-body-decay mechanism, the $K \pi$ pair DAs were introduced, which include the final-state interactions between the $K \pi$ pair in the resonant region. Both the resonant and nonresonant contributions are described by the time-like form factors $F_{K \pi}(s)$, which are parameterized by using the relativistic Breit-Wigner formula for the $\mathrm{P}$-wave resonance $K^*$. Under the narrow width approximation relation and the isospin conservation, the branching ratios for the two-body decays $B_c^+ \to K^{*+} h$ can be related with those of the considered quasi-two-body decays, so it provides us a new way to study these quasi-two-body $B_c$ decays in the three-body framework. We found that the
branching ratios are consistent with those calculated under the two-body framework. It supports the PQCD approach to exclusive hadronic $B_c$ meson decays. While for the direct CP violation, there exists significant difference between the three-body and
the two-body framework. Under the former with the kinematics fixed, the direct CP violation $\mathcal{A}_{CP}$ is just a number, while under the latter the direct CP asymmetry is a differential distribution, which has strong dependence on the $K\pi$ invariant mass $\omega$. It is more convenient to compare with the Dalitz-plot analysis of $A_{CP}$  provided by experiments. The $A_{CP}$ obtained in the two-body framework corresponds to that in the three-body framework with the $K\pi$ invariant mass $\omega$ being fixed to $K^*$ pole mass. Usually, the integration of the $A_{CP}$ differential distribution over the invariant mass $\omega$ under the three-body framework is different with that obtained under the two-body framework, which is usually overestimated or underestimated. It is indeed that compared with the direct CP violations for the decays $B_c \to \ K^{*} D_{(s)}$ obtained in these two frameworks under the PQCD approach, the results from the three-body framework are moderated by the finite width of the $K^*$ resonance, and become comparable with those calculated using other theoretical approaches, such as the Salpeter method and the RCQM. It indicates that it is more appropriate to study the quasi-two-body  $B_c$ meson  decays in the three-body framework than the two-body framework. These results can be tested by the future experiments.

We also researched the annihilation amplitude contributions to the pure annihilation decay modes
$B_c \to \ K^{*} h \to \ K\pi h $ with $h = K, \pi, \eta, \eta'$ in the three-body framework, and found there exists significant
difference for annihilation amplitudes between the $B$ (heavy-light system) and $B_c$ (heavy-heavy system) decays through comparing the branching ratios of the decays $B^+\to \bar K^{*0}K^+, K^{*+}\bar K^0$  and  $B^+_c\to \bar K^{*0}K^+, K^{*+}\bar K^0$. If such point can be clarified by the future experiments, it is helpful to further improve our understanding about the annihilation contributions. Furthermore, among these considered pure annihilation decays, the decay $B^+_c\to  K^{*+}\bar K^0$ has the largest branching ratio and near $10^{-6}$, which is possible observed by the future LHCb experiments.

\section*{Acknowledgment}
We thank Prof. Hsiang-nan Li for valuable discussions. This work is partly supported by the National Natural Science
Foundation of China under Grant No. 11347030, by the Program of
Science and Technology Innovation Talents in Universities of Henan
Province 14HASTIT037, Natural Science Foundation of Henan
Province under Grant No. 232300420116.

{\centering\section{Appendix A: Some relevant functions}}

The explicit expressions of the hard functions $h_i$ with $i=(a, \cdots, h)$ which are obtained from the Fourier transform of the hard kernels are given as
\begin{equation}
\begin{aligned}
h_i\left(\alpha, \beta, b_1, b_2\right) & =h_1\left(\beta, b_2\right) \times h_2\left(\alpha, b_1, b_2\right), \\
h_1\left(\beta, b_2\right) & = \begin{cases}K_0\left(\sqrt{\beta} b_2\right), & \beta>0 ;\\
K_0\left(i \sqrt{-\beta} b_2\right), & \beta<0;\end{cases} \\
h_2\left(\alpha, b_1, b_2\right) & = \begin{cases}\theta\left(b_2-b_1\right) I_0\left(\sqrt{\alpha} b_1\right) K_0\left(\sqrt{\alpha} b_2\right)+\left(b_1 \leftrightarrow b_2\right), & \alpha>0 ; \\
\theta\left(b_2-b_1\right) I_0\left(\sqrt{-\alpha} b_1\right) K_0\left(i \sqrt{-\alpha} b_2\right)+\left(b_1 \leftrightarrow b_2\right), & \alpha<0 ;\end{cases}
\end{aligned}
\end{equation}
where $K_0$ and $I_0$ are modified Bessel functions with $K(ix)=\frac{\pi}{2}(-N_0(x)+iJ_0(x))$ and $J_0$ is Bessel function.
The hard scales $t_i$ are chosen as the maximum of the virtuality of the interal momentum transition in the hard amplitudes and listed as following
\begin{equation}
\begin{aligned}
t_a & =\max \left\{m_{B_c} \sqrt{\left|\alpha_a\right|}, m_{B_c} \sqrt{\left|\beta_a\right|}, 1 / b_3, 1 / b_1\right\}, && t_b=\max \left\{m_{B_c} \sqrt{\left|\alpha_b\right|}, m_{B_c} \sqrt{\left|\beta_b\right|}, 1 / b_1, 1 / b_3\right\}, \\
t_c & =\max \left\{m_{B_c} \sqrt{\left|\alpha_c\right|}, m_{B_c} \sqrt{\left|\beta_c\right|}, 1 / b_1, 1 / b\right\}, && t_d=\max \left\{m_{B_c} \sqrt{\left|\alpha_d\right|}, m_{B_c} \sqrt{\left|\beta_d\right|}, 1 / b_1, 1 / b\right\}, \\
t_e & =\max \left\{m_{B_c}\sqrt{\left|\alpha_e\right|}, m_{B_c} \sqrt{\left|\beta_e\right|}, 1 / b_3, 1 / b_1\right\}, && t_f=\max \left\{m_{B_c} \sqrt{\left|\alpha_f\right|}, m_{B_c} \sqrt{\left|\beta_f\right|}, 1 / b_3, 1 / b_1\right\},\\
t_g & =\max \left\{m_{B_c} \sqrt{\left|\alpha_g\right|}, m_B \sqrt{\left|\beta_g\right|}, 1 / b_3, 1 / b\right\}, && t_h=\max \left\{m_{B_c} \sqrt{\left|\alpha_h\right|}, m_{B_c} \sqrt{\left|\beta_h\right|}, 1 / b, 1 / b_3\right\},
\end{aligned}
\end{equation}
where
\begin{equation}
\begin{aligned}
& \alpha_a=r_b^2+\left(1-r^2\right)\left[(\eta-1) x_3-\eta\right], &&\beta_a=\left(r^2-x_B\right)\left[(1-\eta)\left(x_3-1\right)+x_B\right] , \\
& \alpha_b=\left(r^2-x_B\right)\left(x_B+\eta-1\right), &&\beta_b=\beta_a, \\
& \alpha_c=\beta_a, &&\beta_c=\left[1-x_B-\left(1-r^2\right) z\right]\left[(1-\eta) x_3+x_B-1\right] , \\
& \alpha_d=\beta_a, &&\beta_d=\left[(1-z) r^2-x_B+z\right]\left[(1-\eta)\left(x_3-1\right)+x_B\right], \\
& \alpha_e=(1-\eta)\left(r^2-1\right) x_3 z, &&\beta_e=r_b^2-\left[\left(1-r^2\right) z+x_B-1\right]\left[(1-\eta) x_3+x_B-1\right], \\
& \alpha_f=\alpha_e, &&\beta_f=r_c^2-\left[\left(r^2-1\right) z+x_B\right]\left[(\eta-1) x_3+x_B\right], \\
& \alpha_g=\left(1-r^2\right)\left[(\eta-1) x_3-\eta\right], &&\beta_g=\alpha_e, \\
& \alpha_h=r_c^2+(1-\eta)\left[r^2(z-1)-z\right], &&\beta_h=\alpha_e .\\
&
\end{aligned}
\end{equation}
The Sudakov factors can be written as
\begin{equation}
\begin{aligned}
S_{a b}(t)= & s\left(\frac{m_{B_c}}{\sqrt{2}} x_B, b_1\right)+s\left(\frac{m_{B_c}}{\sqrt{2}} x_3, b_1\right)+\frac{5}{3} \int_{1 / b_1}^t \frac{d \mu}{\mu} \gamma_q(\mu)+2 \int_{1 / b_3}^t \frac{d \mu}{\mu} \gamma_q(\mu), \\
S_{c d}(t)= & s\left(\frac{m_{B_c}}{\sqrt{2}} x_B, b_1\right)+s\left(\frac{m_{B_c}}{\sqrt{2}} z, b\right)+s\left(\frac{m_{B_c}}{\sqrt{2}} (1-z), b\right)+s\left(\frac{m_{B_c}}{\sqrt{2}} x_3, b_1\right) \\
& +\frac{11}{3} \int_{1 / b_1}^t \frac{d \mu}{\mu} \gamma_q(\mu)+2 \int_{1 / b}^t \frac{d \mu}{\mu} \gamma_q(\mu), \\
S_{ef}(t)= & s\left(\frac{m_{B_c}}{\sqrt{2}} x_B, b_1\right)+s\left(\frac{m_{B_c}}{\sqrt{2}} z, b_3\right)+s\left(\frac{m_{B_c}}{\sqrt{2}} (1-z), b_3\right)+s\left(\frac{m_{B_c}}{\sqrt{2}} x_3, b_3\right) \\
& +\frac{5}{3} \int_{1 / b_1}^t \frac{d \mu}{\mu} \gamma_q(\mu)+4 \int_{1 / b_3}^t \frac{d \mu}{\mu} \gamma_q(\mu),\\
S_{gh}(t)= & s\left(\frac{m_{B_c}}{\sqrt{2}} z, b\right)+s\left(\frac{m_{B_c}}{\sqrt{2}} (1-z), b\right)+s\left(\frac{m_{B_c}}{\sqrt{2}} x_3, b_3\right) \\
& +2 \int_{1 / b}^t \frac{d \mu}{\mu} \gamma_q(\mu)+2 \int_{1 / b_3}^t \frac{d \mu}{\mu} \gamma_q(\mu).
\end{aligned}
\end{equation}
As we know, the double logarithms $\alpha_sln^2x$ produced by the radiative corrections are not
small expansion parameters when the end point region is important, in order to improve the
perturbative expansion, the threshold resummation of these logarithms to all order is needed, which
leads to a quark jet function
\be
S_t(x)=\frac{2^{1+2c}\Gamma(3/2+c)}{\sqrt{\pi}\Gamma(1+c)}[x(1-x)]^c,
\en
with $c=0.3$. It is effective to smear the end point singularity with a momentum fraction $x\to0$.


\begin{thebibliography}{99}
\bibitem{lhcb1}
R. Aaij \textit{et al.} [LHCb],
Phys. Rev. D \textbf{94}, 091102 (2016)
[arXiv:1612.07421 [hep-ex]].
\bibitem{lhcb2}
R. Aaij \textit{et al.} [LHCb],
Phys. Rev. D \textbf{95}, 032005 (2017)
[arXiv:1612.07421 [hep-ex]].
\bibitem{lhcb3}
R. Aaij \textit{et al.} [LHCb],
Phys. Lett. B \textbf{759}, 313 (2016)
[arXiv:1603.07037 [hep-ex]].
\bibitem{lhcb4}
R. Aaij \textit{et al.} [LHCb], JHEP \textbf{01}, 065 (2022) [arXiv:2111.03001 [hep-ex]].
\bibitem{Bhat}
B.~Bhattacharya, M.~Gronau and J.~L.~Rosner,
Phys. Lett. B \textbf{726}, 337 (2013)
[arXiv:1306.2625 [hep-ph]].
\bibitem{Gronau}
M.~Gronau,
Phys. Lett. B \textbf{727}, 136 (2013)
[arXiv:1308.3448 [hep-ph]].
\bibitem{xu}
D.~Xu, G.~N.~Li and X.~G.~He, Phys. Lett. B \textbf{728}, 579 (2014)
[arXiv:1311.3714 [hep-ph]].
\bibitem{Gronau2}
M.~Gronau and J.~L.~Rosner, Phys. Rev. D \textbf{72}, 094031 (2005)
[arXiv:hep-ph/0509155 [hep-ph]].
\bibitem{Engelhard}
G.~Engelhard and G.~Raz,
Phys. Rev. D \textbf{72}, 114017 (2005)
[arXiv:hep-ph/0508046 [hep-ph]].
\bibitem{Imbeault}
M.~Imbeault and D.~London,
Phys. Rev. D \textbf{84}, 056002 (2011)
[arXiv:1106.2511 [hep-ph]].
\bibitem{He}
X.~G.~He, G.~N.~Li and D.~Xu,
Phys. Rev. D \textbf{91}, 014029 (2015)
[arXiv:1410.0476 [hep-ph]].
\bibitem{Zhou}
S.~H.~Zhou, R.~H.~Li, Z.~Y.~Wei and C.~D.~Lu,
Phys. Rev. D \textbf{104}, 116012 (2021)
[arXiv:2107.11079 [hep-ph]].
\bibitem{Krankl}
S.~Kr\"ankl, T.~Mannel and J.~Virto,
Nucl. Phys. B \textbf{899}, 247-264 (2015)
[arXiv:1505.04111 [hep-ph]].
\bibitem{Cheng}
H.~Y.~Cheng, C.~K.~Chua and Z.~Q.~Zhang,
Phys. Rev. D \textbf{94}, 094015 (2016)
[arXiv:1607.08313 [hep-ph]].
\bibitem{Li}
Y.~Li,
Phys. Rev. D \textbf{89}, 094007 (2014)
[arXiv:1402.6052 [hep-ph]].
\bibitem{Cheng2007}
H.~Y.~Cheng, C.~K.~Chua and A.~Soni,
Phys. Rev. D \textbf{76}, 094006 (2007)
[arXiv:0704.1049 [hep-ph]].
\bibitem{Klein}
R.~Klein, T.~Mannel, J.~Virto and K.~K.~Vos,
JHEP \textbf{10}, 117 (2017)
[arXiv:1708.02047 [hep-ph]].
\bibitem{zhao}
Y. C. Zhao, Z. Q. Zhang, Z. Y. Zhang, Z. J. Sun and Q. B. Meng, Chin. Phys. C {\bf47}, 073104 (2023) [arXiv:2304.13286 [hep-ph]].
\bibitem{zhang}
Z. Q. Zhang, Y. C. Zhao, Z. L. Guan, Z. J. Sun, Z. Y. Zhang and K. Y. He, Chin. Phys. C {\bf46}, 123105 (2022) [arXiv:2207.02043 [hep-ph]].
\bibitem{zhang2}
Z.~Q.~Zhang and H. x. Guo,
Eur. Phys. J. C {\bf79}, 59 (2019)
[arXiv:1812.11372 [hep-ph]].
\bibitem{Liy}
Y.~Li, W.~F.~Wang, A.~J.~Ma and Z.~J.~Xiao,
Eur. Phys. J. C \textbf{79}, 37 (2019)
[arXiv:1809.09816 [hep-ph]].
\bibitem{maaj}
A. J. Ma, Y. Li and Z. J. Xiao, Nucl. Phys. B {\bf926}, 584 (2018) [arXiv:1710.00327 [hep-ph]].
\bibitem{Chen:2004az}
C.~H.~Chen and H.~n.~Li, Phys. Rev. D \textbf{70}, 054006 (2004) [arXiv:hep-ph/0404097].
\bibitem{Chen:2002th}
C.~H.~Chen and H.~n.~Li,
Phys. Lett. B \textbf{561}, 258 (2003)
[arXiv:hep-ph/0209043].
\bibitem{Liu:2018kuo}
X.~Liu, H.~n.~Li and Z.~J.~Xiao,
Phys. Rev. D \textbf{97}, 113001 (2018)
[arXiv:1801.06145 [hep-ph]].
\bibitem{kuri}
T. Kurimoto, H. n. Li and A. I. Sanda, Phys. Rev. D {\bf67}, 054028 (2003) [arXiv:hep-ph/0210289].
\bibitem{rhli}
R. H. Li, C. D. Lu and Z. Hao, Phys. Rev. D {\bf78}, 014018 (2008) [arXiv:0803.1073 [hep-ph]].
\bibitem{Ma:2017kec}
A.~J.~Ma, Y.~Li, W.~F.~Wang and Z.~J.~Xiao,
Phys. Rev. D \textbf{96}, 093011 (2017)
[arXiv:1708.01889 [hep-ph]].
\bibitem{Ma:2019qlm}
A.~J.~Ma, W.~F.~Wang, Y.~Li and Z.~J.~Xiao, Eur. Phys. J. C \textbf{79}, 539 (2019)
[arXiv:1901.03956 [hep-ph]].
\bibitem{aaij}
R. Aaij \textit{et al.} [LHCb Collaboration], Phys. Rev. D {\bf92}, 012012 (2015) [arXiv:1505.01505 [hep-ex]].
\bibitem{rbw2}
R. Aaij \textit{et al.} [LHCb Collaboration], Phys. Rev. D {\bf94}, 072001 (2016) [arXiv:1608.01289 [hep-ex]].
\bibitem{rbw3}
R. Aaij \textit{et al.} [LHCb Collaboration], Phys. Rev. D {\bf91}, 092002 (2015) [arXiv:1503.02995 [hep-ex]].
\bibitem{heff}
G. Buchalla, A. J. Buras and M. E. Lautenbacher, Rev. Mod. Phys. {\bf68}, 1125 (1996) [arXiv:hep-ph/9512380].
\bibitem{feld}
T. Feldmann, P. Kroll, and B. Stech, Phys. Rev. D {\bf58}, 114006 (1998) [arXiv:hep-ph/9802409].
\bibitem{pdg}
R. L. Workman \textit{et al.} [Particle Data Group], Review of Particle
Physics, PTEP {\bf2022}, 083C01 (2022).
\bibitem{Rui:2011qc}
Z.~Rui, Z.~T.~Zou and C.~D.~Lu,
Phys. Rev. D \textbf{86}, 074008 (2012)
[arXiv:1112.1257 [hep-ph]].
\bibitem{Liu:2009qa}
X.~Liu, Z.~J.~Xiao and C.~D.~Lu,
Phys. Rev. D \textbf{81}, 014022 (2010)
[arXiv:0912.1163 [hep-ph]].
\bibitem{Liu:1997hr}
J.~F.~Liu and K.~T.~Chao,
Phys. Rev. D \textbf{56}, 4133-4145 (1997).
\bibitem{run3}
M. Cepeda, S. Gori, P. Ilten, M. Kado, F. Riva, R. Abdul Khalek, A. Aboubrahim, J. Alimena, S. Alioli and
A. Alves, et al., CERN Yellow Rep. Monogr. 7, 221-584 (2019) [arXiv:1902.00134 [hep-ph]].
\bibitem{choi}
H. M. Choi and C. R. Ji, Phys. Rev. D \textbf{80}, 114003 (2009) [arXiv:0909.5028 [hep-ph]].
\bibitem{hffu}
H. F. Fu, Y. Jiang, C. S. Kim and G. L. Wang, JHEP {\bf1106}, 015 (2011) [arXiv:1102.5399 [hep-ph]].
\bibitem{genon}
 S. D. Genon, J. He, E. Kou and P. Robbe, Phys. Rev. D {\bf80}, 114031 (2009) [arXiv:0907.2256[hep-ph]].
\bibitem{chai}
J. Chai, S. Cheng, Y. H. Ju, D. C. Yang, C. D. Lu and Z. J. Xiao, Chin. Phys. C {\bf46}, 123103 (2022) [arXiv:2207.04190 [hep-ph]].
\bibitem{chua}
H. Y. Cheng and C. K. Chua, Phys. Rev. D {\bf80}, 114008 (2008) [arXiv:0909.5229 [hep-ph]].
\bibitem{wangw}
W. Wang, Y. M. Wang, D. S. Yang and C. D. Lu, Phys. Rev. D {\bf78}, 034011 (2008) [arXiv:0801.3123 [hep-ph]].
\end{thebibliography}
\end{document}